\documentclass[12pt]{article}

\usepackage[margin = 1in]{geometry}
\usepackage{amssymb}
\usepackage{amsmath}
\usepackage{amsthm}
\usepackage{graphicx}
\usepackage{caption}
\usepackage{natbib}
\usepackage[colorlinks = true, citecolor = blue]{hyperref}
\usepackage[svgnames]{xcolor}
\usepackage[ruled,vlined,boxed]{algorithm2e}
\usepackage{stix}
\usepackage{setspace}

\newcommand{\ccal}{\mathcal{C}}

\newcommand{\RR}{\mathbb{R}}

\newcommand{\BF}{\mathbf{F}}
\newcommand{\BG}{\mathbf{G}}

\newcommand{\BI}{\mathbf{I}}

\newcommand{\BK}{\mathbf{K}}

\newcommand{\BP}{\mathbf{P}}

\newcommand{\BS}{\mathbf{S}}

\newcommand{\BV}{\mathbf{V}}
\newcommand{\BW}{\mathbf{W}}
\newcommand{\BX}{\mathbf{X}}
\newcommand{\BY}{\mathbf{Y}}

\newcommand{\Be}{\mathbf{e}}

\newcommand{\Bm}{\mathbf{m}}

\newcommand{\Bp}{\mathbf{p}}

\newcommand{\Bs}{\mathbf{s}}

\newcommand{\Bu}{\mathbf{u}}
\newcommand{\Bv}{\mathbf{v}}
\newcommand{\Bw}{\mathbf{w}}
\newcommand{\Bx}{\mathbf{x}}
\newcommand{\By}{\mathbf{y}}
\newcommand{\Bz}{\mathbf{z}}

\newcommand{\Bmu}{\boldsymbol{\mu}}
\newcommand{\Btheta}{\boldsymbol{\theta}}
\newcommand{\Beta}{\boldsymbol{\eta}}
\newcommand{\Bepsilon}{\boldsymbol{\varepsilon}}

\newcommand{\BLambda}{\boldsymbol{\Lambda}}
\newcommand{\BSigma}{\boldsymbol{\Sigma}}
\newcommand{\BOmega}{\boldsymbol{\Omega}}
\newcommand{\BPsi}{\boldsymbol{\Psi}}
\newcommand{\BGamma}{\boldsymbol{\Gamma}}
\newcommand{\Bgamma}{\boldsymbol{\gamma}}
\newcommand{\BPhi}{\boldsymbol{\Phi}}

\newcommand{\Bphi}{\boldsymbol{\phi}}

\newcommand{\Bzero}{\mathbf{0}}

\newcommand{\tr}{{\scriptscriptstyle\mathsf{T}}}

\newcommand{\argmin}[1]{\underset{#1}{\arg\min}}
\newcommand{\iid}{\overset{\textrm{iid}}{\sim}}
\newcommand{\indep}{\overset{\textrm{indep}}{\sim}}

\newcommand{\dd}{\textrm{d}}
\newcommand{\com}{,\,}
\newcommand{\given}{\,|\,}

\newcommand{\state}{\Bs}

\begin{document}



\title{The projected dynamic linear model for time series on the sphere}

\author{
John Zito\thanks{Assistant Research Professor, Department of Statistical Science, Duke University. Corresponding author: \texttt{john.zito@duke.edu}. Replication files are available in the \texttt{GitHub} repository \href{https://github.com/johnczito/ProjectedDLM}{\texttt{johnczito/ProjectedDLM}}. Zito gratefully acknowledges financial support from the National Science Foundation Graduate Research Fellowship Program under grant number 1842494, and from the Ken Kennedy Institute Computer Science and Engineering Enhancement Fellowship, funded by the Rice Oil and Gas HPC Conference. He also thanks Drs.\ Eduardo García-Portugués, Christophe Ley, and Thomas Verdebout for kind and helpful correspondence.}
\and
Daniel R.\ Kowal
\thanks{Associate Professor, Department of Statistics and Data Science, Cornell University and Associate Professor, Department of Statistics, Rice University. Kowal gratefully acknowledges financial support from the National Science Foundation under grant number SES-2214726.}
}

\maketitle

\begin{abstract}

Time series on the unit $n$-sphere arise in directional statistics, compositional data analysis, and many scientific fields. There are few models for such data, and the ones that exist suffer from several limitations: they are often computationally challenging to fit, many of them apply only to the circular case of $n=2$, and they are usually based on families of distributions that are not flexible enough to capture the complexities observed in real data. Furthermore, there is little work on Bayesian methods for spherical time series. To address these shortcomings, we propose a state space model based on the projected normal distribution that can be applied to spherical time series of arbitrary dimension. We describe how to perform fully Bayesian offline inference for this model using a simple and efficient Gibbs sampling algorithm, and we develop a Rao-Blackwellized particle filter to perform online inference for streaming data. In analyses of wind direction and energy market time series, we show that the proposed model outperforms competitors in terms of point, set, and density forecasting.




\end{abstract}


\section{Introduction}

\begin{figure}
    \centering
    \includegraphics[width = \textwidth]{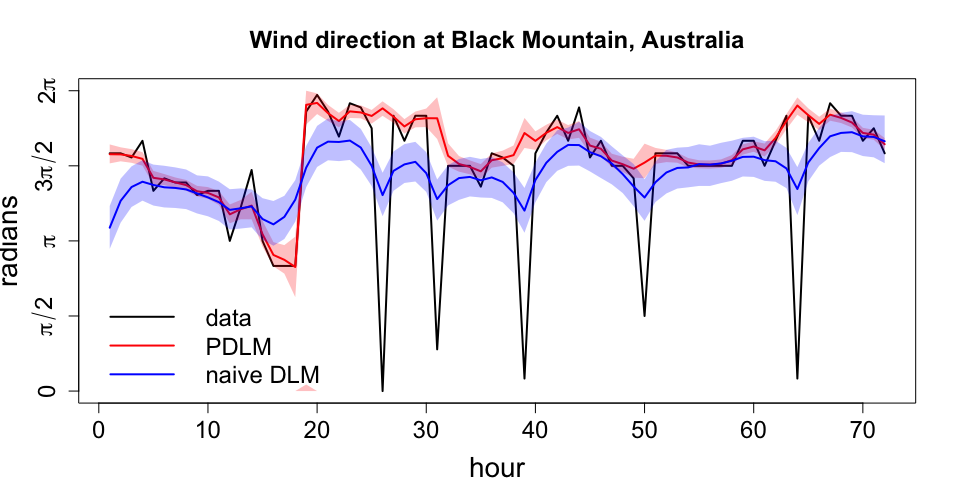}
    \caption{Trend estimates (posterior median and 50\% credible band) from a non-circular model (\ref{eq:dlm_meas}, \ref{eq:dlm_state}) in blue and our circular model (\ref{eq:measurement1}, \ref{eq:transition}, \ref{eq:initialization1}) in red.}
    \label{fig:results}
\end{figure}

Time series that take values on the unit $n$-sphere $S^{n-1}=\{\Bu\in\RR^n:||\Bu||_2=1\}$ arise in many fields. The field of directional statistics studies measurements about the orientation or direction-of-travel of an object, with data recorded as points on $S^1$ or $S^2$ \citep{pg2021test}. Weather stations across the globe, for example, collect high frequency time series on the direction-of-arrival of the wind (Figure~\ref{fig:results}), which is important for monitoring the spread of wildfires or air pollution \citep{gcg2013serra, gbcgp2014serra}. Alternatively,  compositional data analysis studies measurements on the standard simplex $C^{n-1}=\{\Bp\in\RR^{n}:\sum_{i=1}^np_i=1\com p_i\geq0\}$, which can represent proportions of a total such as the market shares of firms in an industry or the fraction of total electricity produced by different energy sources (Figure~\ref{fig:energy}). The simplex-structure of these data poses many challenges, but one promising approach is to apply a square root transformation to compositional data, thus turning it into unit vector data that can be analyzed using spherical methods \citep{sw2011jrssb}.




Because $S^{n-1}\subseteq \RR^n$ is a non-Euclidean manifold, naively applying familiar linear, Gaussian methods to spherical time series can generate distorted and misleading results. To see this, consider the simplest instance of spherical time series: the circular case with $n=2$. In this case, $S^1$ is the unit circle in the plane, so a unit vector $\Bu_t\in S^1$ can  be represented as an angle $a_t\in[0\com 2\pi)$, and the correspondence between the two is given by $\Bu_t=[\cos a_t\,\sin a_t]^\tr$ and $a_t=\text{atan2}(\Bu_t)\text{ mod }2\pi$. Figure~\ref{fig:results} displays a classic time series from \cite{fisher1993book} of the hourly direction-of-arrival of the wind at Black Mountain in Australia. The wind in this example is generally blowing from a south easterly direction, and occasionally passes counterclockwise across the $0/2\pi$ break on the unit circle. On a plot this phenomenon presents as abrupt spikes in the time series of angles. If we naively apply a standard method like a local-level dynamic linear model to these data, we obtain a (smoothed) trend estimate (blue) that overfits to the spikes and overstates the uncertainty because the linear, Gaussian model does not understand the ``wrapped'' structure of the circle. By comparison, applying the proposed  projected dynamic linear model (Section~\ref{sec:model}) that \emph{does} understand this structure, we obtain more satisfactory trend estimates and uncertainty bands (red). 

\subsection{Previous work}

We prioritize the following properties for modeling spherical time series:
\begin{itemize}
    \item \textbf{data coherence}: naive linear, Gaussian methods generate distorted results for spherical data (Figure~\ref{fig:results}). It is necessary to apply methods that take into account the special structure of the sphere by, for instance, making use of probability distributions supported on the sphere;
    
    \item \textbf{generalize to the hyperspherical case}: circular data ($n=2$) is an important special case, but models should also generalize to the 
   spherical ($n=3$) and hyperspherical ($n>3$) cases;
   
    \item \textbf{generate full predictive distributions}: we prioritize flexible predictive distributions, rather than just point forecasts, that acknowledge as many sources of uncertainty as possible. This is most commonly achieved using Bayesian inference via the posterior predictive distribution;
    \item \textbf{sequential inference for streaming data}: in the time series setting, observations are often streaming by in real-time and potentially at high frequency. We wish to recursively update our inferences and predictions in an efficient way that adapts to incoming data;
    \item \textbf{exact inference}: statistical computations should be ``exact'' in the sense that estimates, predictions, or posterior distributions can be approximated arbitrarily well in the limit of some algorithm setting that the user controls, such as the number of Monte Carlo samples;
    \item \textbf{jointly estimate static and dynamic unobservables}: for state space models, there are often unknown static parameters. These should be estimated jointly with the dynamic latent states. 
\end{itemize}
We review the models proposed in the literature---each of which has major shortcomings in several of these areas (see Table~\ref{tab:compare}).

\begin{table}[h]
    \centering
    \begin{tabular}{l | llllll}\hline
     & coherent & general & predictive & sequential & exact & parameters \\\hline 
     naive (V)AR & No & Yes & Yes & Yes & Yes & Yes \\
     naive DLM & No & Yes & Yes & partial & Yes & partial \\
     WAR & Yes & No & Yes & No & Yes & Yes \\
     LAR & Yes & No & Yes & No & Yes & Yes \\
     WN-SSM & Yes & No & Yes & Yes & No & No \\
     vMF-SSM & Yes & Yes & Yes & Yes & No & No \\
     (D)SAR & Yes & Yes & No & No & Yes & Yes \\
     PDLM & Yes & Yes & Yes & partial & Yes & partial \\\hline
\end{tabular}
    \caption{The proposed \emph{projected dynamic linear model} (PDLM) fully or partially satisfies all of our desired properties of a time series model for spherical data.}
    \label{tab:compare}
\end{table}

Naive linear, Gaussian methods like the autoregression (AR) or the dynamic linear model (DLM) are simple and convenient to implement and understand, but their lack of data coherence makes them fundamentally inappropriate. The first data coherent time series models were developed by \cite{craig1988thesis}, \cite{breckling1989book}, and \cite{fl1994jrssb} with the aim of extending the ARMA framework to circular data. This work generated the wrapped autoregression (WAR) and the linked autoregression (LAR). Recently these models have been extended by \cite{hmsz2006ees}, \cite{abmp2015jsip}, \cite{hp2023jtsa}, and \cite{hhpt2024joe} to allow regime-switching and hidden Markov structure. These models focus solely on the case of circular ($n=2$) time series, with no straightforward avenue toward generalizing them to the hyperspherical case. The models are all based on either the von Mises (vM) distribution or the wrapped normal distribution, which are families of distributions on the unit circle whose densities are restricted to being symmetric and unimodal. Finally, these models are developed from a frequentist point of view, and estimation can be computationally challenging, with no standard software implementations.

In parallel to these developments, various state space models have been proposed, which are summarized in \cite{kgh2016ieeeaesm}. These state space models have a certain recursive Bayesian interpretation, but they do not admit analytically tractable recursions for filtering or smoothing, so inference is approximated using unscented transformations and deterministic sampling. As with the classical models, the state space models are limited to the circular case, and they only make use of the vM or wrapped normal distribution. This seriously limits the flexibility of the (approximate) forecasting distributions that the models generate. Furthermore, there is little or no practical guidance on how to estimate the static parameters that govern these models. There is software for these models in the form of the \texttt{Matlab} toolbox \texttt{libDirectional} \citep{kgpdhhs2019jss}.

Beyond this modest literature on time series analysis for circular data, there is much less work on the general spherical case ($n\geq 3)$. \cite{zm2023joe} recently defined a general spherical autoregressive model. They derive Yule-Walker-type estimators for the model's parameters, derive their asymptotic properties, and apply the model to compositional data problems. However, this approach only provides point predictions and not full distributional forecasts.  \cite{kgh2016ieeespl} generalized  \cite{kgh2016ieeeaesm} by using the von Mises-Fisher (vMF) distribution to define the measurement and state transition distributions of a state space model, thereby allowing the model to scale to arbitrary values of $n$. However, this model retains all of the limitations of the circular ones on which it is based: the vMF distribution is not very flexible, inference in the model is only approximate, and the static parameters are not estimated.

\subsection{The projected normal distribution}

\begin{figure}
    \centering
    \includegraphics[scale = 0.175]{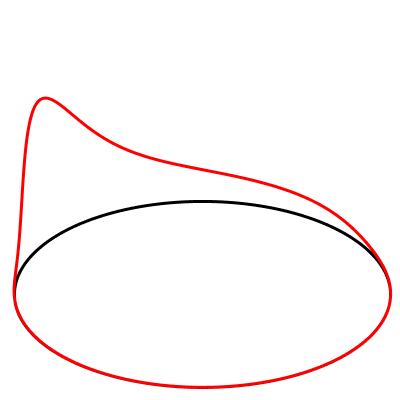}
    \includegraphics[scale = 0.2]{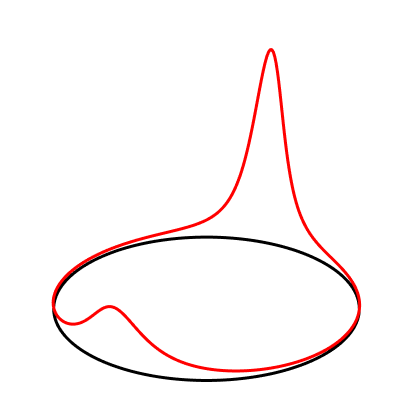}
    \includegraphics[scale = 0.2]{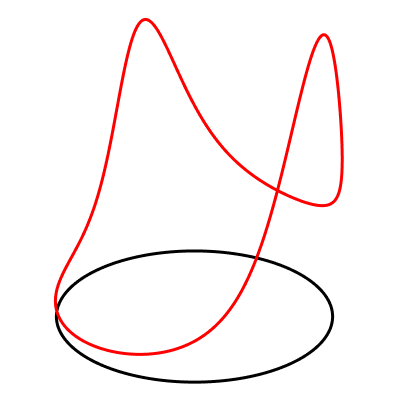}
    \includegraphics[scale = 0.2]{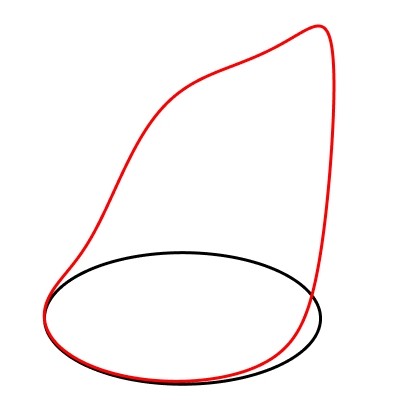}
    \includegraphics[scale = 0.2]{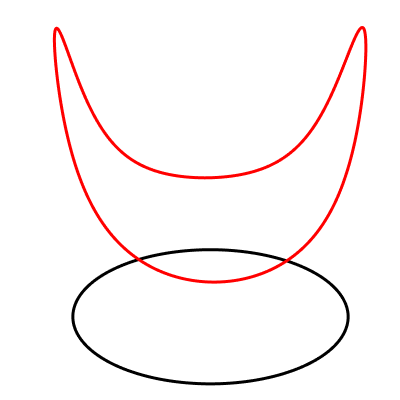}
    \caption{The projected normal density can exhibit asymmetry, bimodality, and antipodality.}
    \label{fig:pn}
\end{figure}

A recurring and critical shortcoming of existing work is the reliance on the vM(F) and wrapped normal distributions, which have restricted shapes and may not be flexible enough for the demands of real data. This has motivated a surge of recent interest in the projected normal distribution for non-dynamic data \citep{pml1998jasa,ng2005jas,wg2013sm,hbv2017ba}. If $\Bx\sim\textrm{N}_n(\Bmu\com\BSigma)$, then $\Bu=\Bx/||\Bx||_2\in S^{n-1}$ has the \emph{projected normal distribution}, denoted $\Bu\sim\textrm{PN}_n(\Bmu\com\BSigma)$. One of the advantages of this parametric family is that it encompasses much more flexible distributional shapes than its competitors. The formula for the projected normal density on the unit circle is given in \cite{wg2013sm}, and Figure~\ref{fig:pn} displays examples of the shapes it can exhibit. Whereas the vM(F) and wrapped normals densities are only ever symmetric and unimodal, the projected normal density can be asymmetric, bimodal, and antipodal. This means that a fully Bayesian model based on the projected normal distribution has the potential to generate much more flexible predictive distributions.

Another advantage of this class of distributions is that the connection to the normal distribution enables convenient statistical computation. Considering the static case for a moment, if we observe $\Bu\sim\textrm{PN}_n(\Bmu\com\BSigma)$, we can think of this as a partially observed Gaussian random vector $\Bx=r\Bu\sim\textrm{N}_n(\Bmu\com\BSigma)$, where the length $r>0$ is an unobserved latent variable. This representation enables classical inference for the parameters $\Bmu$ and $\BSigma$ via an EM algorithm  \citep{pml1998jasa} or Bayesian inference via a Gibbs sampler that alternates between sampling the parameters and sampling the length  \citep{ng2005jas}. In this way, the introduction of the latent length functions as a data augmentation in the same spirit as \cite{ac1993jasa} for Bayesian probit regression or \cite{dr1994jrssb} for Gaussian mixture models.

\subsection{Our contribution}

In this paper, we combine the computational tractability and distributional flexibility of the projected normal distribution with the tools of dynamic linear models to introduce a state space model that can be applied to spherical time series of arbitrary dimension (Section~\ref{sec:model}). The model can be represented as a partially-observed linear, Gaussian state space model, which allows us to perform fully Bayesian inference for all of the model's unobservables (both dynamic latent states and static parameters) using a simple and efficient Gibbs sampler based on the familiar Kalman simulation smoother. Leveraging this unique model tractability, we introduce a Rao-Blackwellized particle filter to perform online inference for streaming data (Section~\ref{sec:rbpf}). As such, our model is the first with the potential to realize all of the desired properties that previous models have been unable to balance (Table~\ref{tab:compare}). To put this model to its fullest use, we provide the first full enumeration of how to produce and evaluate point, set, and density predictions on the hypersphere (Section~\ref{sec:forecasting}). These tools are applicable beyond the proposed model and apply for any Bayesian model that generates a predictive distribution on the hypersphere. In applications to wind direction and energy market composition data, we find that our model outperforms others in terms of real-time, out-of-sample probabilistic forecasting (Section~\ref{sec:applications}). Supplementary material includes additional computing details (Appendix~\ref{app:trend}), validation of the sampling algorithm (Appendix~\ref{app:gir}), and verification of parameter recovery for simulated data (Appendix~\ref{app:consistency}).

\section{The projected dynamic linear model}\label{sec:model}

To study the dynamics of a time series $\Bu_{1:T}$ on the unit $n$-sphere, we propose the \emph{projected dynamic linear model} (PDLM):
\begin{align}
    \Bu_t\given\Bs_t
    &\indep 
    \textrm{PN}_n\big(\BF_t\state_t\com \BSigma\big)
    \label{eq:measurement1}\\
    \state_t
    &=
    \BG
    \state_{t-1}
    +
    \Beta_t
    ,
    &&
    \Beta_t
    \iid
    \textrm{N}_p(\Bzero\com\BW)
    \label{eq:transition}
    \\
    \state_0
    &\sim 
\textrm{N}_p\big(\bar{\state}_{0|0}\com\BP_0\big)
    \label{eq:initialization1}
\end{align}
for $t=1,\ldots,T$. 
This is a state space model where the observed unit vectors $\Bu_t\in S^{n-1}$ depend on the latent time series $\Bs_t\in\RR^p$ which we must estimate. The observed and unobserved time series are linked by the measurement distribution in (\ref{eq:measurement1}), where the projected normal family ensures distributional flexibility, tractable computation, and our ability to handle spherical observations of any dimension $n$. The $n\times p$ matrices $\BF_t$ are fixed and known, and we can manipulate them to produce different model structures. If we take $p = n$ and all $\BF_t=\BI_n$, then the PDLM becomes a signal-in-noise, ``local level'' type model for spherical time series (see Figure~\ref{fig:results}). If we take $p=nm$ and set $\BF_t=\BI_n\otimes \Bx_t^\tr$ for some vector of observed covariates $\Bx_t\in\RR^m$, then the PDLM becomes a spherical-on-linear regression model like those studied in \cite{pml1998jasa} and \cite{ng2011sm}, only now $\Bs_t$ can be interpreted as a vector of time-varying regression coefficients.

The model is governed by the static parameters $\Btheta =(\BSigma\com\BG\com\BW)$, which are typically unknown and must be estimated. The latent state $\Bs_t$ evolves according to the VAR(1) given in (\ref{eq:transition}), and $\BG$ and $\BW$ determine these dynamics. In some cases we may choose to fix these hyperparameters (like setting $\BG=\BI_p$ for random walk state evolution), but in general we can estimate them. For convenience we choose a (conditionally) conjugate matrix normal inverse Wishart prior:
\begin{equation}
    \BG\com\BW\sim \textrm{MNIW}_{p\com p}\big(\nu_0\com\BPsi_0\com\overline{\BG}_0\com\BOmega^{-1}_0\big).\label{eq:gw}
\end{equation}
We fix the prior hyperparameters to impose the desired amount of shrinkage. Typically, we truncate the prior on $\BG$ so that it always has eigenvalues inside the unit circle, which enforces stationarity of $\Bs_t$ and thus $\Bu_t$. The parameter $\BSigma$ influences the shape of the measurement distribution, and in general it is not identified because $\textrm{PN}_n(\Bmu\com \BSigma)$ and $\textrm{PN}_n(c\Bmu\com c^2\BSigma)$ are the same distribution for any $c>0$. To ensure identifiability, we impose the following reparametrization and priors due to \cite{hbv2017ba}:
\begin{align}
    \BSigma&=
    \begin{bmatrix}
    \BGamma + \Bgamma\Bgamma^\tr & \Bgamma \\
    \Bgamma^\tr & 1
    \end{bmatrix},
    \quad
    \begin{matrix}
    \BGamma\sim\textrm{IW}_{n-1}(d_0\com \BPhi_0)\\
    \Bgamma\sim\textrm{N}_{n-1}(\bar{\Bgamma}_0\com\BLambda_0).
    \end{matrix}\label{eq:sig}
\end{align}

As in the static case, the key to inference in the PDLM is to augment the ``incomplete data'' specification in (\ref{eq:measurement1}, \ref{eq:transition}, \ref{eq:initialization1}) with a latent length variable $r_t$ for each observation. This gives the ``complete data'' or ``data-augmented'' version of the model:
\begin{align}
    r_t\Bu_t
    &=
    \BF_t\state_t
    +
    \Bepsilon_t
    ,
    \quad
    \Bepsilon_t
    \iid
    \textrm{N}_n
    \left(
    \Bzero
    \com 
    \BSigma
    \right)
    \label{eq:measurement2}.
\end{align}
Together with (\ref{eq:transition}, \ref{eq:initialization1}), this describes a linear, Gaussian state space model, but the response variable $r_t\Bu_t\in\RR^n$ is only partially observed; we observe $\Bu_t\in S^{n-1}$ but not $r_t>0$. So the goal of Bayesian inference is to access the full posterior $p(\state_{0:T}\com r_{1:T}\com \Btheta\given \Bu_{1:T})$. Once we have draws from this distribution, the latent lengths are of no independent interest, and they can be discarded so that we are left with a sample from the original PDLM posterior $p(\state_{0:T}\com \Btheta\given \Bu_{1:T})$.

To access the data-augmented posterior, we use the Markov chain Monte Carlo (MCMC) sampler described in Algorithm~\ref{alg:pdlm_gibbs_sampler}. The algorithm is a Gibbs sampler that requires us to iteratively simulate the conditional posterior of each set of unobservables: $\Bs_{0:T}$, $\BGamma$, $\Bgamma$, $\BG$, $\BW$, $r_{1:T}$. Fortunately, this can mostly be done analytically:
\begin{itemize}
    \item Conditional on both $r_t$ and $\Bu_t$, we are given ``data'' $\By_t=r_t\Bu_t$ and thus model in (\ref{eq:measurement2}, \ref{eq:transition}, \ref{eq:initialization1}) 
    becomes a \emph{fully-observed} linear, Gaussian state space model. This means we can access $p(\state_{0:t}\given r_{1:t}\com \Btheta\com \Bu_{1:t})$ exactly using the Kalman simulation smoother;
    \item Conditional on $\Bu_t$, $r_t$, and $\Bs_t$,  the quantities $\By_t=r_t\Bu_t$, $\Bmu_t=\BF_t\Bs_t$, and $\Bz_t=\By_t-\Bmu_t=\begin{bmatrix}\Bz_{-n,t} & z_{n,t}\end{bmatrix}^\tr$ are all known. From (\ref{eq:measurement2}) and (\ref{eq:sig}), we see that $\Bz_t\iid\textrm{N}_n(\Bzero\com\BSigma)$ and $\Bz_{-n,t}\given z_{n,t}\indep\textrm{N}_{n-1}(\Bgamma z_{n,t}\com \BGamma)$. If we condition on $\Bgamma$, then $\Bz_{-n,t}-\Bgamma z_{n,t}$ is a fully observed, iid sample from a multivariate normal with conjugate inverse Wishart prior on $\BGamma$, and so $p(\BGamma\given\state_{0:T}\com r_{1:T}\com\Bgamma\com\BG\com\BW\com \Bu_{1:T})$ is also inverse Wishart;
    \item If instead we condition on $\BGamma$, then $\Bz_{-n,t}\given z_{n,t}\indep\textrm{N}_{n-1}(\Bgamma z_{n,t}\com \BGamma)$ describes a vector-on-scalar regression with known covariance matrix. Given the normal prior on $\Bgamma$, the conditional posterior $p(\Bgamma\given\state_{0:T}\com r_{1:T}\com\BGamma\com\BG\com\BW\com \Bu_{1:T})$ is also normal with parameters given, for instance, in \cite{koop2003book};
    \item $p(\BG\com\BW\given\state_{0:T}\com r_{1:T}\com\BGamma\com\Bgamma\com \Bu_{1:T})$ is the posterior of a VAR(1) with a conjugate MNIW prior, where the $\Bs_{0:T}$ are treated as the data, and thus is also MNIW (e.g., \citealp{karlsson2013chapter}).
\end{itemize}
The only conditional posterior that is intractable is that of $r_{1:T}$, but this can be accessed using the slice sampler from \cite{hbv2017ba} that is reproduced in Algorithm~\ref{alg:slice}. With that, a slice-within-Gibbs sampler is given in Algorithm~\ref{alg:pdlm_gibbs_sampler} that enables us to conduct fully Bayesian inference for all of the unobservables in the PDLM. In Appendix~\ref{app:gir}, we validate this algorithm with the method of \cite{geweke2004jasa}, and in Appendix~\ref{app:consistency} we perform simulations showing that the posterior provides consistent inference for the static parameters $\Btheta$.




\begin{algorithm}
\caption{Gibbs sampling steps for $p(\state_{0:T}\com r_{1:T}\com \BGamma\com\Bgamma\com\BG\com\BW\given \Bu_{1:T})$}
\For{$p(\state_{0:T}\given r_{1:T}\com \BGamma\com\Bgamma\com\BG\com\BW\com \Bu_{1:T})$}{
Construct pseudo-data $\By_t=r_t\Bu_t$ and then draw $\state_{0:T}$ from the posterior of the linear, Gaussian state space model in (\ref{eq:measurement2}, \ref{eq:transition}, \ref{eq:initialization1}) using an implementation of the Kalman simulation smoother: \cite{dk2002bka}, \cite{cj2009ijmmno}, etc. 
}
\For{$p(\BGamma\given\state_{0:T}\com r_{1:T}\com\Bgamma\com\BG\com\BW\com \Bu_{1:T})$}{
Let $\By_t=r_t\Bu_t$, $\Bmu_t=\BF_t\Bs_t$, $\Bz_t=\By_t-\Bmu_t=\begin{bmatrix}\Bz^\tr_{-n,t} & z_{n,t}\end{bmatrix}^\tr$, and $\Be_t=\Bz_{-n,t}-\Bgamma z_{n,t}$\;

Draw $\BGamma\sim\textrm{IW}_{n-1}(d_T\com \BPhi_T)$, where $d_T=d_0+T$ and $\BPhi_T=\BPhi_0+\sum\limits_{t=1}^T\Be_t\Be_t^\tr$.
}
\For{$p(\Bgamma\given\state_{0:T}\com r_{1:T}\com\BGamma\com\BG\com\BW\com \Bu_{1:T})$}{
Construct pseudo-data
$$
\Bw=\begin{bmatrix}
    \Bz_{-n,1}^\tr&\Bz_{-n,2}^\tr&\cdots &\Bz_{-n,T}^\tr
\end{bmatrix}^\tr
,\quad 
\Bv=\begin{bmatrix}
    z_{n,1}&z_{n,2}&\cdots &z_{n,T}
\end{bmatrix}^\tr
,\quad 
\BV=\Bv\otimes \BI_{n-1}.
$$

Compute
\begin{align*}
\BLambda_T&=\left[\BLambda_0^{-1}+\BV^\tr\left(\BI_T\otimes\BGamma^{-1}\right)\BV\right]^{-1}\\
    \bar{\Bgamma}_T&=\BLambda_T\left[\BLambda_0^{-1}\bar{\Bgamma}_0+\BV^\tr\left(\BI_T\otimes\BGamma^{-1}\right)\Bw\right].
\end{align*}

Draw $\Bgamma\sim\textrm{N}_{n-1}\left(\bar{\Bgamma}_T\com\BLambda_T\right)$.
}
\For{$p(\BG\com\BW\given\state_{0:T}\com r_{1:T}\com\BGamma\com\Bgamma\com \Bu_{1:T})$}{
Construct pseudo-data 
$\BY = \begin{bmatrix}
\state_1 & \state_2 & \cdots & \state_T
\end{bmatrix}^\tr$ and $\BX = \begin{bmatrix}
\state_0 & \state_1 & \cdots & \state_{T-1}
\end{bmatrix}^\tr
$.

Compute 
\begin{align*}
    \nu_T&=\nu_0+T\\
    \BOmega_T&=\BX^\tr\BX+\BOmega_0\\
    \overline{\BG}_T&=\BOmega_T^{-1}(\BX^\tr\BY+\BOmega_0\overline{\BG}_0)\\
    \BPsi_T&=\BPsi_0+\left(\BY-\BX\overline{\BG}_T\right)^\tr\left(\BY-\BX\overline{\BG}_T\right)+\left(\overline{\BG}_T-\overline{\BG}_0\right)^\tr\BOmega_0\left(\overline{\BG}_T-\overline{\BG}_0\right).
\end{align*}

Draw $\BG\com\BW\sim \textrm{MNIW}_{p\com p}(\nu_T\com\BPsi_T\com\overline{\BG}_T\com\BOmega^{-1}_T)$ until $\BG$ has eigenvalues in the unit circle.
}
\For{$p(r_{1:T}\given\state_{0:T}\com\BGamma\com\Bgamma\com\BG\com\BW\com \Bu_{1:T})$}{
(\textbf{parallel}) \For{$t = 1\com 2\com ...\com T$}{
Draw $r_t^{(m)}$ using Algorithm~\ref{alg:slice} with $r^{(\text{old})}=r_t^{(m-1)}$, $\Bu=\Bu_t$, $\Bm=\BF_t\Bs_t$, $\BS=\BSigma$.
}
}
\label{alg:pdlm_gibbs_sampler}
\end{algorithm}

\begin{algorithm}
\caption{Slice sampling step for $p(r\given \Bu\com\Bm\com\BS)$ when $\Bu\in S^{n-1}$ and $r\Bu\sim\textrm{N}_n(\Bm\com\BS)$}
\label{alg:slice}
\textbf{Input}: $r^{(\text{old})}$, $\Bu$, $\Bm$, $\BS$\;
Compute $a=\Bu^\tr\BS^{-1}\Bu$ and $b=\Bu^\tr\BS^{-1}\Bm$\;
Draw independent $v\sim\textrm{Unif}\left(0\com \exp\left[-\frac{a}{2}\left(r^{(\text{old})}-\frac{b}{a}\right)^2\right]\right)$ and $u\sim\textrm{Unif}(0\com1)$\;
Compute 
$
c
=
\frac{b}{a} 
+
\max
\left\{ 
-\frac{b}{a}
\com 
-
\sqrt{
\frac{-2\ln v}{a}
}
\right\}
$
and 
$
d
= 
\frac{b}{a} 
+
\sqrt{
\frac{-2\ln v}{a}
}
$\;
\textbf{Return}: $r^{(\text{new})}=\left[\left(d^n-c^n\right)u+c^n\right]^{1/n}$.
\end{algorithm}



\section{Sequential inference}\label{sec:rbpf}

Algorithm~\ref{alg:pdlm_gibbs_sampler} describes how to perform ``batch'' or ``offline'' inference for the PDLM. In this environment, all of the historical data $\Bu_{1:T}$ that we wish to analyze are available at once, and we can use our Gibbs sampler to generate full information estimates of all of the PDLM unobservables: both static parameters and dynamic latent states. But because we are modeling time series data, we must also consider the ``streaming'' or ``online'' environment where the observations $\Bu_t$ are arriving one-after-another in real-time, and we wish to recursively update our posterior approximation to keep up with the new information. Our Gibbs sampler can be used for this, but it will be computationally inefficient. When generating a sample from the time $t$ posterior, naive MCMC does not provide a means of recycling the old sample from the time $t-1$ posterior and adjusting it to reflect the influence of the new observation. 
This ignores the fact that time-adjacent posteriors will typically be quite similar, since they differ by only one observation in their conditioning set. Furthermore, if the data are streaming by at high frequency, there may not be enough time between observations to run the MCMC algorithm long enough.





To overcome these issues and perform online inference for streaming data, we design a particle filter for the PDLM. This algorithm uses importance sampling to recursively construct a weighted Monte Carlo sample from the sequence of data-tempered posteriors. Unlike a naive MCMC approach, this algorithm \emph{does} recycle and adjust the draws from the previous period in order to avoid reprocessing the entire history of data from scratch. This makes the algorithm much more efficient in terms of clock time per-period than the MCMC algorithm when faced with streaming data. Ideally, our online algorithm would recursively traverse the sequence of full posteriors $p(\Bs_{0:t}\com r_{1:t}\com \Btheta\given \Bu_{1:t})$ just as our MCMC algorithm does in the offline setting. Unfortunately, online inference for both the states and the parameters of a state space model is notoriously difficult in general \citep{kdsmc2015statsci}, and we will not confront this issue here. In what follows we will assume that the static parameters $\Btheta$ are fixed and known, and we henceforth drop them from the notation. Furthermore, while in theory our algorithm targets the sequence of full smoothing distributions $p(\Bs_{0:t}\com r_{1:t}\given \Bu_{1:t})$, we only apply it to the filtering problem of recursively approximating the marginal posterior $p(\state_t\com r_t\given \Btheta\com \Bu_{1:t})$. Because of the Markovian structure of the model, this is sufficient for accessing the model's forecast distribution in real-time.

As with the Gibbs sampler, the inclusion of $r_t$ makes the filtering problem challenging, but because of the model's unique tractability, we will not have to sample it jointly with $\Bs_t$ in the usual way. To see this, consider this marginal-conditional decomposition of the full posterior:
\begin{equation}
    p(\state_{0:t}\com r_{1:t}\given \Bu_{1:t})
=
p(\state_{0:t}\given r_{1:t}\com \Bu_{1:t})p(r_{1:t}\given \Bu_{1:t}).
\end{equation}
$p(\state_{0:t}\com r_{1:t}\given \Bu_{1:t})$ is not tractable, but as we discussed in Section~\ref{sec:model}, we can access $p(\state_{0:t}\given r_{1:t}\com \Bu_{1:t})$ exactly via the Kalman filter and smoother. The well-known filtering recursions are restated in Table~\ref{tab:kalman} for convenience. In order, then, to access $p(\state_{0:t}\com r_{1:t}\given \Bu_{1:t})$, it suffices to access the marginal posterior of the latent lengths $p(r_{1:t}\given \Bu_{1:t})$. Once we have a sample $\{r_{1:t}^{(m)}\}\sim p(r_{1:t}\given \Bu_{1:t})$, we can perfectly simulate $\state_{0:t}^{(m)}\sim p(\state_{0:t}\given r_{1:t}^{(m)}\com \Bu_{1:t})$ to produce draws from the full posterior. This sampling scheme amounts to a Rao-Blackwellization that can improve the Monte Carlo efficiency of the posterior approximation by significantly reducing the dimension of the problem \citep{rr2021isr}. Instead of applying Monte Carlo methods to the $(p+1)$-dimensional problem of sampling $\Bs_{t}$ and $r_{t}$ jointly, we need only apply them to the one dimensional problem of sampling $r_t$. The rest of the posterior is handled exactly, with no added Monte Carlo error.


We will use a particle filter to sequentially traverse the sequence of posteriors $p(r_{1:t}\given \Bu_{1:t})$ as new observations $\Bu_t$ arrive. In order to do this, we need to calculate the unnormalized posterior density. We know that
    \begin{equation}\label{eq:target1}
        p(r_{1:t}\given \Bu_{1:t})\propto p(r_{1:t}\com \Bu_{1:t}),
    \end{equation}
    and in general we know that
    \begin{equation}\label{eq:target2}
        p(r_{1:t}\com \Bu_{1:t})=\prod_{k=1}^tp(r_k\com \Bu_k\given r_{1:k-1}\com \Bu_{1:k-1}).
    \end{equation}
    From Table~\ref{tab:kalman}, we know that
    \begin{equation}
        p(r_k\Bu_k\given r_{1:k-1}\com \Bu_{1:k-1})=\textrm{N}_n(r_k\Bu_k;\,\bar{\By}_{k|k-1}\com \BOmega_{k|k-1}).
    \end{equation}
    If we apply the change of variables $r_k\Bu_k\mapsto (r_k\com \Bu_k)$ from Euclidean to polar coordinates, we get
    \begin{equation}\label{eq:target3}
         p(r_k\com \Bu_k\given r_{1:k-1}\com \Bu_{1:k-1})
    =
    r_k^{n-1}
    \textrm{N}_n(r_k\Bu_k;\,\bar{\By}_{k|k-1}\com \BOmega_{k|k-1})q(\Bu_k),
    \end{equation}
    where $q(\cdot)$ is the part of the Jacobian that does not depend on $r_k$. Combining (\ref{eq:target1}), (\ref{eq:target2}), and (\ref{eq:target3}), we see that 
    the posterior kernel is
    \begin{equation}\label{eq:target4}
         p(r_{1:t}\given \Bu_{1:t})\propto
    \prod_{k=1}^t
    r_k^{n-1}
    \textrm{N}_n(r_k\Bu_k;\,\bar{\By}_{k|k-1}\com \BOmega_{k|k-1}).
    \end{equation}

\begin{table}
    \centering
    \begin{tabular}{l l l}\hline 
        Object & Distribution & Details\\ \hline
        \\
        $\state_t\given (r\Bu)_{1:t-1}$ & $\textrm{N}_p\left(\bar{\state}_{t|t-1}\com\BP_{t|t-1}\right)$ & $\bar{\state}_{t|t-1}=\BG \bar{\state}_{t-1|t-1}$\\
                                  & & $\BP_{t|t-1}=\BG\BP_{t-1|t-1}\BG^\tr+\BW$\\
                                  \\
        $r_t\Bu_t\given (r\Bu)_{1:t-1}$ & $\textrm{N}_n\left(\bar{\By}_{t|t-1}\com\BOmega_{t|t-1}\right)$ & $\bar{\By}_{t|t-1}=\BF_t\bar{\state}_{t|t-1}$\\
                                  & & $\BOmega_{t|t-1}=\BF_t\BP_{t|t-1}\BF_t^\tr+\BSigma$\\
                                  \\
        $\state_t\given (r\Bu)_{1:t}$   & $\textrm{N}_p\left(\bar{\state}_{t|t}\com\BP_{t|t}\right)$ & $\bar{\state}_{t|t}=\bar{\state}_{t|t-1}+\BP_{t|t-1}\BF_t^\tr\BOmega_{t|t-1}^{-1}(r_t\Bu_t-\bar{\By}_{t|t-1})$\\
                                  & & $\BP_{t|t}=\BP_{t|t-1}-\BP_{t|t-1}\BF_t^\tr\BOmega_{t|t-1}^{-1}\BF_t\BP_{t|t-1}$\\ 
                                  \\\hline
    \end{tabular}
    \caption{Kalman filter recursions for updating $\BK_t$ during the RBPF.}
    \label{tab:kalman}
\end{table}

This distribution is not tractable, but we now have a formula for evaluating its unnormalized density pointwise, and that is sufficient to target the sequence with a particle filter. Because our target kernel is the result of having marginalized out the $\state_t$, this will be a Rao-Blackwellized particle filter (RBPF) in the sense of \cite{dfmr2000puai}. The goal of a particle filter is to recursively construct a weighted sample $\{r_{1:t}^{(m)}\com W_t^{(m)}\}_{m=1}^M$ that is adapted to $p(r_{1:t}\given \Bu_{1:T})$, and as \cite{dj2011chapter} explain, at the foundation of the method is sequential importance sampling. Given our target kernel
    \begin{equation}
        k(r_{1:t})=\prod_{k=1}^t
    r_k^{n-1}
    \textrm{N}_n(r_k\Bu_k;\,\bar{\By}_{k|k-1}\com \BOmega_{k|k-1}),
    \end{equation}
    and given a sequential importance proposal density
    \begin{equation}
        g(r_{1:t})=g_1(r_1)\prod_{k=2}^tg_k(r_k\given r_{1:k-1}),
    \end{equation}
    we have the following recursion for the importance weights:
    \begin{align*}
        W_t&=\frac{k(r_{1:t})}{ g(r_{1:t})}
        =\frac{k(r_{1:t-1})}{k(r_{1:t-1})}\frac{k(r_{1:t})}{ g(r_{1:t})}
        =\frac{k(r_{1:t-1})}{g(r_{1:t-1})}\frac{k(r_{1:t})}{k(r_{1:t-1})g_t(r_t)}
        =W_{t-1}\frac{k(r_{1:t})}{k(r_{1:t-1})g_t(r_t)}.
    \end{align*}
    Invoking the particular form of $k$, this yields
    \begin{equation}\label{eq:weights}
        W_t=W_{t-1}\frac{r_t^{n-1}\textrm{N}_n(r_t\Bu_t;\,\bar{\By}_{t|t-1}\com \BOmega_{t|t-1})}{g_t(r_t)}.
    \end{equation}

    We see in (\ref{eq:weights}) that in order to compute the importance weights, we need access to $\bar{\By}_{t|t-1}$ and $\BOmega_{t|t-1}$. That is to say, we need access to the Kalman filter output that exactly characterizes the conditional posterior of the state $\Bs_t$. So as we adapt our ``swarm'' $\{r_{1:t}^{(m)}\com W_t^{(m)}\}_{m=1}^M$ of particles from period to period, we must also update the statistics that characterize $p(\state_{t}\given r_{1:t}^{(m)}\com\Bu_{1:t})$. Recalling Table~\ref{tab:kalman}, denote these statistics $\BK_t=\left\{\bar{\state}_{t|t-1}\com \BP_{t|t-1}\com \bar{\By}_{t|t-1}\com\BOmega_{t|t-1}\com\bar{\state}_{t|t}\com\BP_{t|t}\right\}$. We can now attach to each particle its own $\BK_t^{(m)}$ and update them recursively as the algorithm runs. Table~\ref{tab:kalman} describes the map $(\BK_{t-1}\com r_t\com \Bu_t)\mapsto\BK_t$ that updates the statistics from one period to the next.


With these preliminaries sorted, we give the RBPF in Algorithm~\ref{alg:particle_filter}. Using the language of \cite{chopin2004aos}, one recursion proceeds in three phases: correction, selection, and mutation. During the correction phase, we propose new values, update the statistics $\BK_t$, and update the importance weights. A simple choice for $g_t$ is a random walk in logs $\text{LogNormal}(\ln r_{t-1}\com\sigma^2_g)$, where $\sigma^2_g$ is a tuning parameter. In order to prevent the importance weights from degenerating, the selection phase checks whether or not the effective sample size (ESS) of the swarm has dropped below a user-defined threshold $\tau$, and if it has we resample the particles in a multonomial fashion and make the weights uniform. Resampling helps prevent degeneracy in the weights, but at the cost of additional Monte Carlo noise, since the resulting swarm will now contain duplicate values and be less diverse. To combat this, the mutation phase replenishes the particles by ``jittering'' them according to an MCMC algorithm that targets $p(r_t\given r_{1:t-1}\com\Bu_{1:t})$. For this, we  apply the slice sampler in Algorithm~\ref{alg:slice}.

\begin{figure}
    \centering
    \includegraphics[width = 0.45\textwidth]{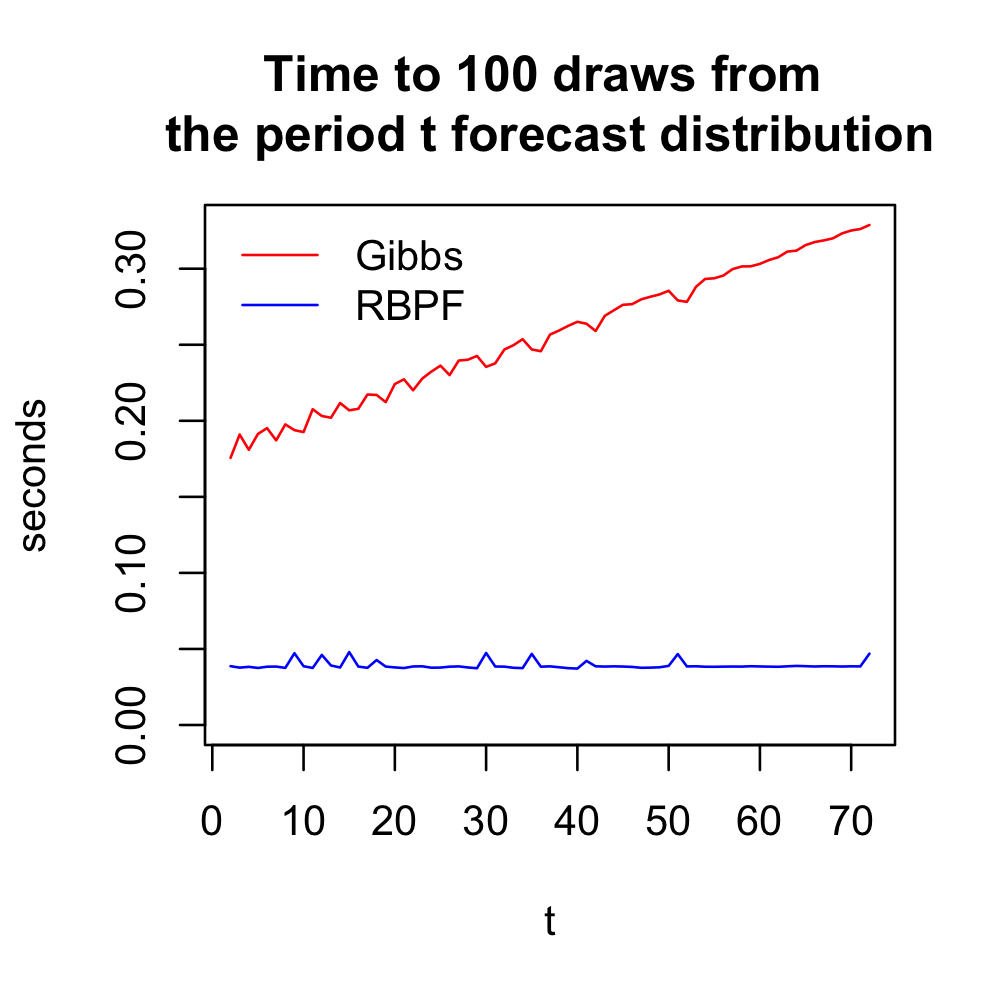}
    \includegraphics[width = 0.45\textwidth]{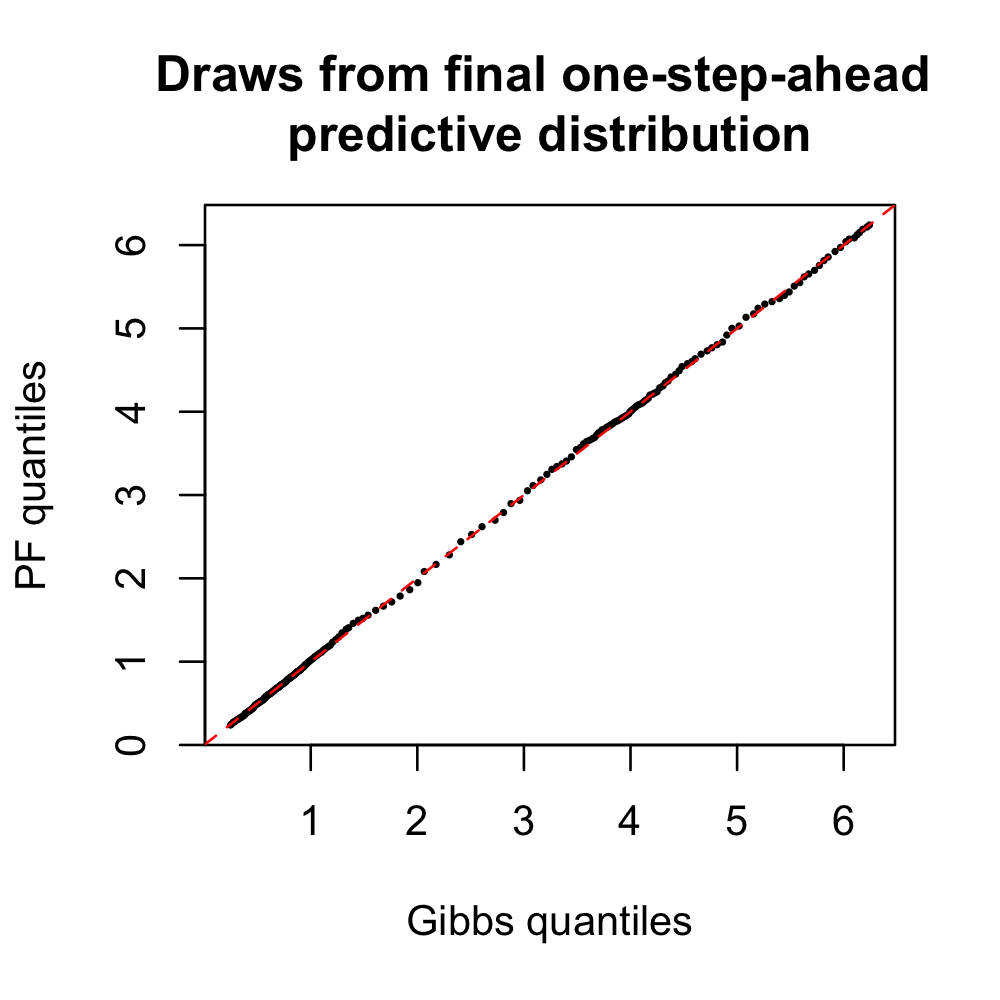}
    \caption{\textbf{Left}: The time in seconds (median over ten repetitions) that it takes to generate 100 draws from the forecast distribution $p(\Bu_t\given \Bu_{1:t-1})$ each period. For the Gibbs sampler this requires reprocessing the entire history of data, so the per-period computing cost scales with the sample size. \textbf{Right}: Q-Q plot comparing the draws generated by the Gibbs sampler (Algorithm~\ref{alg:pdlm_gibbs_sampler}: $M=5000$, burn = 5000, thin = 5) and the RBPF (Algorithm~\ref{alg:particle_filter}: $M=2500$, $\tau = 0.5$, $\sigma^2_g=0.25$). Given the same priors and data (the wind direction series in Figure~\ref{fig:results}), they should give similar output.}
    \label{fig:correct}
\end{figure}

Given the same data and priors, Algorithm~\ref{alg:pdlm_gibbs_sampler} (holding $\Btheta$ fixed and alternating between the first and last steps) and Algorithm~\ref{alg:particle_filter} both produce draws from the same posterior distribution. Indeed, Figure~\ref{fig:correct} (right) is a Q-Q plot comparing the draws from the period $T+1$ predictive distribution that we get when we apply either algorithm to the data in Figure~\ref{fig:results}. Thankfully, they agree in their output. But these two algorithms are best suited to different data environments. Algorithm~\ref{alg:pdlm_gibbs_sampler} is meant for the batch or offline setting where all of the data have arrived and we seek a single good approximation to the time $T$ posterior. Algorithm~\ref{alg:particle_filter} is meant for the streaming or online setting where we wish to recursively update our posterior approximation as new observations arrive period-after-period in real-time. Algorithm~\ref{alg:pdlm_gibbs_sampler} could also be used in this context, but it will be less efficient.


Figure~\ref{fig:correct} (left) displays the time (median over ten repetitions) that each algorithm takes to recursively generate one hundred draws from the forecast distribution $p(\Bu_{t+1}\given \Bu_{1:t})$. We see that the Gibbs sampler handles this task in $O(t)$ time, whereas the particle filter handles it in $O(1)$ time, which is the signal difference between offline and online inference. The MCMC algorithm is offline because it does not ``recycle'' the period $t-1$ posterior approximation when it constructs the period $t$ approximation in the way that the RBPF does. When a new observation arrives, the MCMC user must simply rerun their sampler from scratch in order to incorporate the new information. As the sample size grows observation-by-observation, yet more computations will mount as we re-process the entire history of data each period. In a high frequency streaming environment, there may not be time to rerun the MCMC algorithm each period, especially when the time series history is long. In this case, the RBPF is more suitable.

    


\begin{algorithm}
\caption{Rao-Blackwellized particle filter for $p(r_t\given \BGamma\com\Bgamma\com\BG\com\BW\com \Bu_{1:t})$}
\label{alg:particle_filter}
\textbf{Input}: New observation $\Bu_t$; old particles $\left\{\BK_{t-1}^{(m)}\com r_{t-1}^{(m)}\com W_{t-1}^{(m)}\right\}_{m=1}^M$ adapted to $p(r_{t-1}\given \Bu_{1:t-1})$\;

\For{ (\textbf{Correction}) }{
    
    (\textbf{parallel}) \For{$m=1,\com2\com...\com M$}{
    Propose $\tilde{r}_t^{(m)}\sim g_t(r_t\given r_{t-1}^{(m)}\com...)$\;
    Update $\left(\BK_{t-1}^{(m)}\com \tilde{r}_t^{(m)}\com \Bu_t\right)\mapsto \tilde{\BK}_t^{(m)}$ using Table~\ref{tab:kalman}\;
    Reweight
    $$
    \tilde{W}_t^{(m)}\propto W_{t-1}^{(m)}\frac{\left[\tilde{r}^{(m)}_t\right]^{n-1}\textrm{N}_n\left(\tilde{r}^{(m)}_t\Bu_t;\,\bar{\By}^{(m)}_{t|t-1}\com{\BOmega}^{(m)}_{t|t-1}\right)}{g_t\left(\tilde{r}^{(m)}_t\right)}
    ;
    $$
    }
}    
\For{ (\textbf{Selection}) }{
Compute $\text{ESS}=M/\left(M^{-1}\sum\limits_{m=1}^M\left[\tilde{W}_t^{(m)}\right]^2\right)^{-1}$\;
\For{$m=1,\com2\com...\com M$}{
\eIf{$\text{ESS}<\tau$}{
Draw $(\BK_t^{(m)}\com r_t^{(m)})\sim \{\tilde{\BK}_t^{(j)}\com \tilde{r}_t^{(j)}\com \tilde{W}_t^{(j)}\}_{j=1}^M$\;
Set $W_t^{(m)}=1/M$\;
}{
Set $(\BK_t^{(m)}\com r_t^{(m)}\com W_t^{(m)})=(\tilde{\BK}_t^{(m)}\com \tilde{r}_t^{(m)}\com \tilde{W}_t^{(m)})$\;
}
}
}
    
\For{(\textbf{Mutation})}{
(\textbf{parallel}) \For{$m=1,\com2\com...\com M$}{
    Mutate $r_t^{(m)}$ for $L$ iterations of Algorithm~\ref{alg:slice} with $\Bu=\Bu_t$, $\Bm=\bar{\By}^{(m)}_{t|t-1}$, $\BS=\BOmega_{t|t-1}^{(m)}$\;
    Update $\bar{\state}_{t|t}^{(m)}$ in $\BK^{(m)}_t$ based on the final value of $r_t^{(m)}$.
    }
}

\textbf{Return}: New particles $\left\{\BK_{t}^{(m)}\com r_{t}^{(m)}\com W_{t}^{(m)}\right\}_{m=1}^M$ adapted to $p(r_{t}\given \Bu_{1:t})$.
\end{algorithm}

\section{Probabilistic forecasting with spherical data}\label{sec:forecasting}

In this work we are primarily interested in probabilistic forecasting for spherical time series. That is, we are interested in using data $\Bu_{1:t}\subseteq S^{n-1}$ to produce and evaluate a full forecast distribution $f_{t+1|t}(\Bu)$ on the sphere that incorporates as many sources of uncertainty as possible about the future observation $\Bu_{t+1}$. Once we have a forecast distribution $f_{t+1|t}$, we can also extract from it point forecasts $\hat{\Bu}_{t+1|t}\in S^{n-1}$ and forecast sets $C_{t+1|t}\subseteq S^{n-1}$. To operationalize this, we consider the case where we have Monte Carlo output $\{\tilde{\Bu}^{(j)}_{t+1}\}_{j=1}^J\sim f_{t+1|t}(\Bu)$ from the forecast distribution. In this setting we seek to use these draws to approximate and evaluate point, set, and density predictions on the sphere. Although \cite{wg2014jasa} consider spatial point and density (but not set) prediction for circular data ($n=2$), methods for point, set, and density prediction in the general (hyper)spherical case have yet to be enumerated in the literature.

For our purposes specifically, we operate in a Bayesian framework where we take as our forecast distribution $f_{t+1|t}(\Bu)$ the posterior predictive distribution:
$$p(\Bu_{t+1}\given \Bu_{1:t})=\int p(\Bu_{t+1}\given \Bu_{1:t}\com\Bphi)p(\Bphi\given \Bu_{1:t})\,\dd\Bphi.$$
For the PDLM,  Algorithm~\ref{alg:pdlm_gibbs_sampler} (or \ref{alg:particle_filter}) produces a sample $\{\Bphi^{(j)}\}_{j=1}^J\sim p(\Bphi\given \Bu_{1:t})$ from the full posterior distribution of the dynamic and static unobservables $\Bphi=\{\Bs_{1:t+1}\com r_{1:t}\com\Btheta\}$. For each draw $\Bphi^{(j)}$, we simulate forward the model dynamics  in (\ref{eq:measurement1}) to produce a Monte Carlo sample $\{\tilde{\Bu}^{(j)}_{t+1}\}_{j=1}^J\sim p(\Bu_{t+1}\given \Bu_{1:t})$ from the predictive distribution. So in practice, we apply the methods described below to the specific task of summarizing posterior predictive simulations. But it is important to bear in mind that these techniques are completely generic, and they would apply to Monte Carlo output $\{\tilde{\Bu}^{(j)}_{t+1}\}_{j=1}^J\sim f_{t+1|t}(\Bu)$ coming from any predictive distribution $f_{t+1|t}(\Bu)$ produced by any method. Furthermore, we describe everything in terms of one-period-ahead forecasting for simplicity, but the extension to $h$-periods-ahead is straightforward.

\subsection{Point forecasting}

To generate a point forecast, we can calculate a summary statistic of the draws $\{\tilde{\Bu}^{(j)}_{t+1}\}_{j=1}^J$, such as the sample mean or median direction. We opt for the latter \citep{fisher1985jrssb}, which is given by
\begin{equation}\label{eq:med}
\hat{\Bu}_{t+1|t}
=
\argmin{\Bm\in S^{n-1}}
\,
\frac{1}{J}
\sum\limits_{j=1}^J
\cos^{-1}\big(\Bm^\tr\tilde{\Bu}^{(j)}_{t+1}\big),
\end{equation}
where $d(\Bu\com\Bv)=\cos^{-1}(\Bu^\tr\Bv)$ is the geodesic distance on the hypersphere. We can also use this metric to measure the error between a forecast $\hat{\Bu}$ and a realization $\Bu$. So given a sequence of point forecasts $\hat{\Bu}_{t+1|t}$ and subsequent realizations $\Bu_{t+1}$, the mean spherical forecast error of a method is 
\begin{equation}\label{eq:mce}
\textrm{MSpFE}=\frac{1}{T-t_0-1}\sum\limits_{t=t_0}^{T-1}\cos^{-1}\big(\hat{\Bu}_{t+1|t}^\tr\Bu_{t+1}\big).
\end{equation}
This measures the average quality of the one-step-ahead point forecasts (lower is better).

\subsection{Set forecasting}

\begin{figure}
    \centering
    \includegraphics[width=\textwidth]{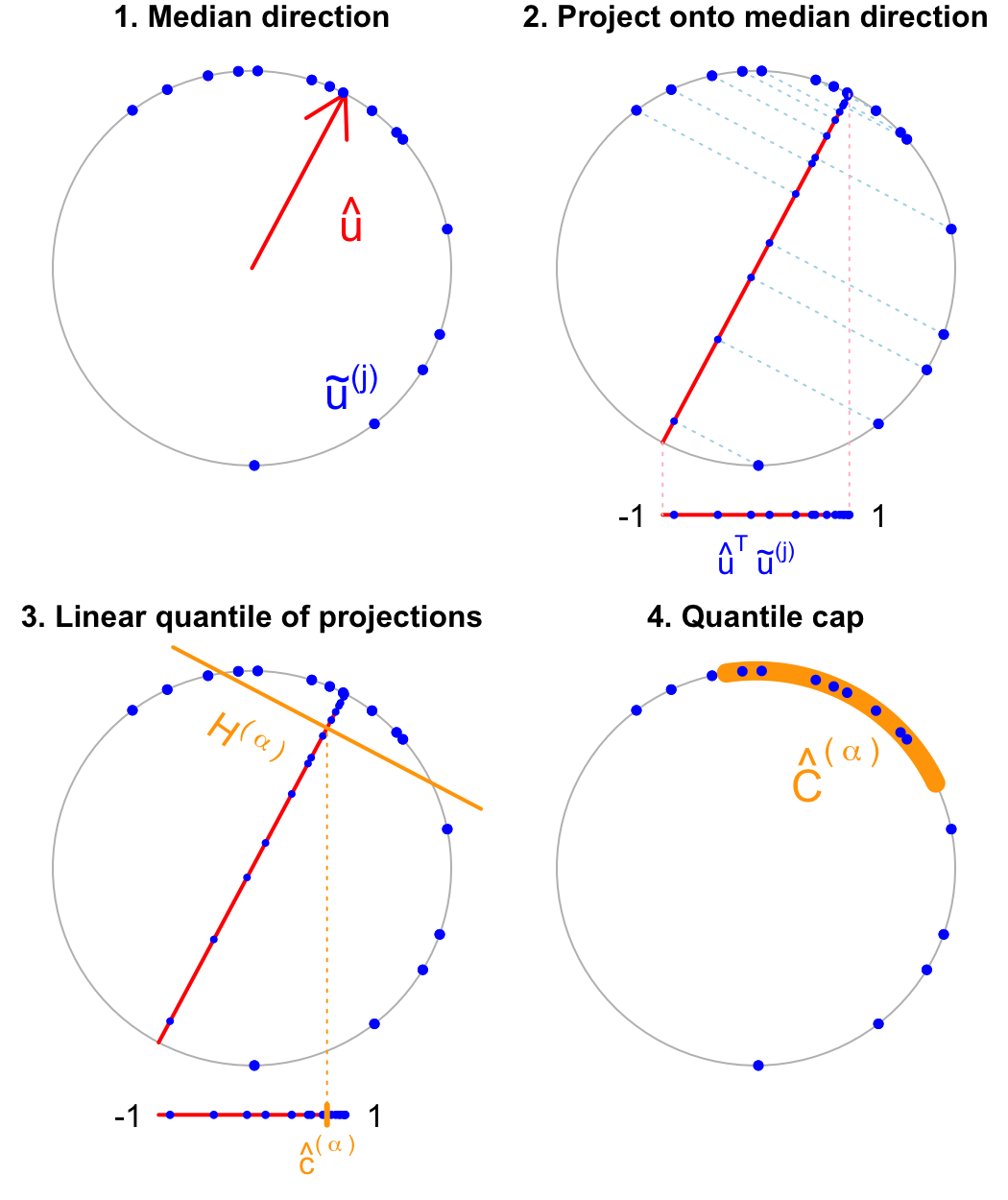}
    \caption{
    Constructing quantile caps. Given a sample $\{\tilde{\Bu}^{(j)}\}_{j=1}^{J}\subseteq S^{1}$ of size $J=16$,
    we (1) compute the median direction $\hat{\Bu}$, (2) project each $\tilde{\Bu}^{(j)}$ onto $\hat{\Bu}$, (3) compute the ordinary, linear $\alpha=0.5$ quantile $\hat{c}^{(\alpha)}$ of the scalar projections $\hat{\Bu}^\tr\tilde{\Bu}^{(j)}$, thereby defining a hyperplane $H^{(\alpha)}$ that bisects $S^1$ into two caps, and (4) select the upper cap $\hat{\ccal}^{(\alpha)}$ as our set forecast. When all is said and done, we see that $\hat{\ccal}^{(\alpha)}$ is centered at $\hat{\Bu}$, and it contains $(1-\alpha)J=0.5\cdot16=8$ of the draws, as it should.
    }
    \label{fig:quantile}
\end{figure}

To generate a set forecast, we consider posterior predictive credible sets $\ccal_{t+1|t}^{(\alpha)}\subseteq S^{n-1}$ satisfying $P(\Bu_{t+1}\in\ccal_{t+1|t}^{(\alpha)}\given\Bu_{1:t})=1-\alpha$. To approximate such a set using our draws $\{\tilde{\Bu}_{t+1}^{(j)}\}$, we calculate the upper quantile cap $\hat{\ccal}_{t+1|t}^{(\alpha)}$ of \cite{lsv2014ejs}, which satisfies $J^{-1}\sum_{j=1}^J1\big(\tilde{\Bu}_{t+1}^{(j)}\in\hat{\ccal}_{t+1|t}^{(\alpha)}\big)\approx 1-\alpha$. In the original context, the quantile cap is used as a descriptive statistic for spherical data. Here, we use it as a Monte Carlo estimate of a posterior summary (the credible set). This calculation proceeds in four steps, which we visualize in Figure~\ref{fig:quantile} for the case of the unit circle $S^1$:
\begin{enumerate}
    \item We start with the sample median direction $\hat{\Bu}_{t+1|t}$ as in (\ref{eq:med}), and construct a credible set centered around it. In this way, $\hat{\ccal}_{t+1|t}^{(\alpha)}$ is a (hyper)spherical analog to a prediction interval of the form $\hat{y}\pm\varepsilon$; 
    \item we project each draw $\tilde{\Bu}_{t+1}^{(j)}$ onto the median direction, which produces the univariate sample $\{\hat{\Bu}_{t+1|t}^\tr\tilde{\Bu}_{t+1}^{(j)}\}\subseteq[-1\com 1]$;
    \item we compute the ordinary sample quantile of the scalar projections, which \cite{lsv2014ejs} call the \emph{projection quantile}: 
    $$
\hat{c}^{(\alpha)}_{t+1|t}
=
\argmin{-1\leq c\leq 1}
\frac{1}{J}\sum\limits_{j=1}^J\rho_\alpha\left(\hat{\Bu}_{t+1|t}^\tr\tilde{\Bu}_{t+1}^{(j)}-c\right),
\quad 
\rho_\alpha(y)=y(\alpha-1_{y\leq 0}).
$$
This defines a hyperplane $H^{(\alpha)}$ that is orthogonal to $\hat{\Bu}_{t+1|t}$ and intersects it at the point $\hat{c}^{(\alpha)}_{t+1|t}\hat{\Bu}_{t+1|t}$. $H^{(\alpha)}$ bisects $S^{n-1}$ into two upper and lower \emph{quantile caps};
    \item we take for our forecast set the upper quantile cap:
    $$
\hat{\ccal}_{t+1|t}^{(\alpha)}
=
\left\{ 
\Bu\in S^{n-1}
:
\hat{\Bu}_{t+1|t}^\tr\Bu
\geq
\hat{c}^{(\alpha)}_{t+1|t}
\right\}.
$$
By construction, $100\times(1-\alpha)\%$ of the posterior predictive draws $\tilde{\Bu}_{t+1}^{(j)}$ have $\hat{\Bu}_{t+1|t}^\tr\tilde{\Bu}_{t+1}^{(j)}
\geq
\hat{c}^{(\alpha)}_{t+1|t}$, thus placing them in $\hat{\ccal}_{t+1|t}^{(\alpha)}$.
\end{enumerate}

Given a sequence of set forecasts $\hat{\ccal}_{t+1|t}^{(\alpha)}$ and subsequent realizations $\Bu_{t+1}$, we can evaluate the performance of the sets by computing their average size and coverage over time:
\[
    \text{size}_\alpha=\frac{1}{T-1-t_0}\sum\limits_{t=t_0}^{T-1}\text{area}(\hat{\ccal}_{t+1|t}^{(\alpha)}), \quad 
    \text{coverage}_\alpha=\frac{1}{T-1-t_0}\sum\limits_{t=t_0}^{T-1}1\big(\Bu_{t+1}\in\hat{\ccal}_{t+1|t}^{(\alpha)}\big).
\]
We seek high-coverage sets of modest size. Checking whether $\Bu_{t+1}\in\hat{\ccal}_{t+1|t}^{(\alpha)}$ is as simple as checking if $\hat{\Bu}_{t+1|t}^\tr\Bu_{t+1}\geq\hat{c}^{(\alpha)}_{t+1|t}$. We compute the size of each $\hat{\ccal}_{t+1|t}^{(\alpha)}$ using the surface area formula in \cite{li2011ajms}.

\subsection{Density forecasting}

The sample $\{\tilde{\Bu}^{(j)}_{t+1}\}_{j=1}^J$ represents a discrete approximation of the forecast distribution, so to score the distribution forecast as a whole, we use a proper kernel scoring rule like in \cite{sz2021acha}. If $f_{t+1|t}(\Bu)$ is a forecast distribution on $S^{n-1}$ and $\Bu_{t+1}$ the subsequent realization, then 
\begin{equation}\label{eq:kernel}
    S\left(f_{t+1|t}\com \Bu_{t+1}\right)=\frac{1}{2}E\left[g(\tilde{\Bu}\com\tilde{\Bu}')\right]-E\left[g(\tilde{\Bu}\com \Bu_{t+1})\right], \quad \tilde{\Bu}\com\tilde{\Bu}'\iid f_{t+1|t}
\end{equation}
is a proper scoring rule so long as the kernel function $g:S^{n-1}\times S^{n-1}\to\RR$ is symmetric positive definite. \cite{gneiting2013bern} studied several such functions on the sphere, and in this work we take the powered exponential kernel $g(\Bu\com\Bv)=\exp(-\cos^{-1}(\Bu^\tr\Bv))$. In this case where $f_{t+1|t}$ is represented by the Monte Carlo sample $\{\tilde{\Bu}^{(j)}_{t+1}\}_{j=1}^J$, we approximate the kernel score in (\ref{eq:kernel}) with 
\begin{equation}
    \hat{S}\left(\{\tilde{\Bu}^{(j)}_{t+1}\}\com \Bu_t\right) = \frac{1}{2}\frac{1}{J^2}\sum\limits_{j=1}^J\sum\limits_{k=1}^Jg(\tilde{\Bu}_{t+1}^{(j)}\com \tilde{\Bu}_{t+1}^{(k)})-\frac{1}{J}\sum\limits_{j=1}^Jg(\tilde{\Bu}_{t+1}^{(j)}\com \Bu_{t+1}).
\end{equation}
To score the entire sequence of forecast distributions, we calculate each method's mean kernel score over time:
\begin{equation}\label{eq:mcrps}
    \textrm{MKS}=\frac{1}{T-1-t_0}\sum\limits_{t=t_0}^{T-1}\hat{S}\big(\{\tilde{\Bu}^{(j)}_{t+1}\}\com \Bu_{t+1}\big).
\end{equation}
The smaller the MKS is, the better the forecast distributions are on average over time.

\section{Competing models}\label{sec:models}

In the forecasting comparisons presented in Section~\ref{sec:applications}, we compare our PDLM to a basket of alternatives. The first is a naive local-level dynamic linear model (DLM) like we applied in Figure~\ref{fig:results}:
\begin{align}
a_t&=s_t+\varepsilon_t,&&\varepsilon_t\iid\textrm{N}(0\com v)\label{eq:dlm_meas}\\
    s_t&=s_{t-1}+\eta_t,&&\eta_t\iid\textrm{N}(0\com w)\label{eq:dlm_state}
\end{align}
with independent normal and inverse gamma priors on $s_0$, $v$, and $w$. This is fit using using a Gibbs sampler supplied in the \texttt{dlm} package in \texttt{R} \citep{petris2010jss}. For circular data, we fit this model to an observed time series of angles $a_t\in[0\com 2\pi)$, but because the model does not incorporate the circular structure of the data, it produces posterior predictive distributions $p(a_{t+1}\given a_{1:t})$ mistakenly supported on the entire real line. To correct for this, we define the posterior predictive functional $\Bu_{t+1}=[\cos a_{t+1}\,\sin a_{t+1}]^\tr\in S^1$ and use $p(\Bu_{t+1}\given a_{1:t})$ to generate forecasts. So as a post-processing step, we forcibly project the ill-supported posterior predictive output to the correct space. As we will see, this naive correction is largely cosmetic, and does not repair the underlying issues with applying a naive linear, Gaussian model to these data.

Next, we consider the local-level state space model proposed by \cite{kgh2016ieeespl}, which we abbreviate vMF-SSM:
        \begin{align}
        \Bu_t\given \Bs_t &\indep \textrm{vMF}_n(\Bs_t\com \kappa)\\
        \Bs_t\given \Bs_{t-1} &\sim \textrm{vMF}_n(
    \Bs_{t-1}\com \eta).
    \end{align}
$\textrm{vMF}_n(...)$ denotes the von Mises-Fisher distribution. Unlike the DLM, this is a proper, data-coherent time series model for spherical data of arbitrary dimension, and in principle it will produce posterior predictive distributions with the correct support. Unfortunately, it does not admit closed-form recursions for filtering and smoothing, and so these operations must be approximated using unscented transformations and deterministic sampling. This is implemented in the \texttt{Matlab} toolbox \texttt{libDirectional} \citep{kgpdhhs2019jss}. The literature does not provide guidance on how to jointly estimate the static parameters $\kappa$ and $\eta$, so we fix $\eta=1$ and estimate $\kappa$ from a pre-sample using the method in \cite{sra2012cs}.

Lastly, we compare to the spherical autoregression (SAR) of \cite{zm2023joe}, which is another data-coherent method that is properly supported on the unit sphere of arbitrary dimension. The SAR assumes $\Bu_t$ is stationary, and models it in terms of its ``spherical difference'' $R_t=\Bu_t\ominus\Bmu$ from a time-invariant Fr\'{e}chet mean $\Bmu$:
\begin{align}
    R_t-\mu_R
    &= 
    \alpha_1(R_{t-1}-\mu_R)
    +
    \cdots 
    +
    \alpha_p(R_{t-p}-\mu_R)
    +
    \varepsilon_t.
\end{align}
Here, $\alpha_l\in \RR$, $\mu_R=E[R_t]$, and $\varepsilon_t$ are iid mean zero. \cite{zm2023joe} also discuss a differenced version of the model (DSAR) where the autoregression applies to the first spherical difference of the data $R_t=\Bu_{t+1}\ominus\Bu_t$, which allows their framework to be applied to certain kinds of nonstationary spherical time series. Yule-Walker type estimators are used for $\alpha_i$ and the  lag order $p$ is selected using time series cross-validation. Given a fitted model, this method produces point predictions $\hat{\Bu}_{t+1|t}$, but the authors do not provide guidance about generating prediction sets or full predictive distributions, so we cannot produce and evaluate these in our comparisons.

\section{Applications}\label{sec:applications}

To evaluate the performance of the PDLM compared to the other models described in Section~\ref{sec:models}, we perform recursive, one-step-ahead probabilistic forecast comparisons. For each $t=t_0\com ...\com T$, we retrain all of our models on the data up to period $t-1$, and then generate the point, interval, and density forecasts for period $t$. We compare the forecasts to the subsequent realizations of $\Bu_{t}$, and we average the forecasting performance over time for each model as described in Section~\ref{sec:forecasting}.

\subsection{Wind direction on Black Mountain, Australia}\label{sec:wind}

Figure~\ref{fig:results} displays the hourly direction-of-arrival of the wind at Black Mountain, Australia recorded in radians \citep{fisher1993book}. This is a time series on the unit circle $S^1$. Table~\ref{tab:results} presents one-hour-ahead forecast metrics averaged over the period $t_0=11$ to the end of the sample $T=72$. We see that the PDLM delivers the best point predictions, the best density predictions, and the smallest set predictions with close-to-nominal coverage. Consistent with Figure~\ref{fig:results}, the naive DLM delivers the worst point predictions by far, but comparable set and density predictions to the otherwise data-coherent vMF-SSM. Among the data-coherent methods, the SAR (whose lag length tuning parameter is reoptimized via cross-validation in every stage of the exercise), delivers the worst point predictions and does not generate set or density predictions.

\begin{table}
    \centering
    \begin{tabular}{l | llll}\hline
     & MSpFE  &  size  & coverage  & MKS  \\\hline 
DLM     & 0.768 & 4.917    & 0.951 & -0.198 \\
SAR     & 0.642 & --    & -- & -- \\
vMF-SSM &0.615& 5.207  &  0.951& -0.203 \\
PDLM    & 0.602 &3.562  &  0.919 &-0.252 \\
     \hline
\end{tabular}
    \caption{One-step-ahead, out-of-sample forecasting metrics for the Black Mountain data, averaged over the period $t_0=11$ to $T = 72$. Forecast sets are 90\% posterior predictive credible sets.}
    \label{tab:results}
\end{table}

\subsection{United States energy market composition}\label{sec:energy}

\begin{figure}
    \centering
    \includegraphics[width = \textwidth]{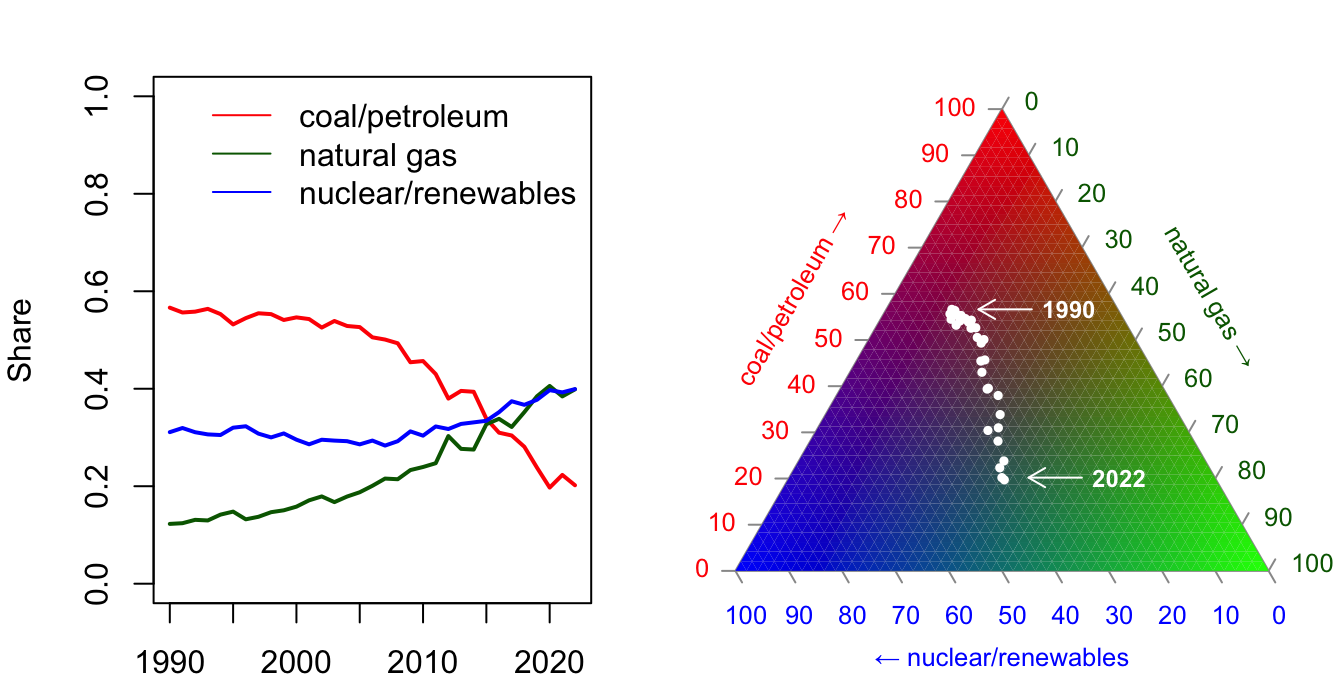}
    \caption{The share of annual US electricity generated by various sources (1990 - 2022).}
    \label{fig:energy}
\end{figure}

Figure~\ref{fig:energy} displays the yearly composition of total US electricity generation from 1990 to 2022\footnote{US Energy Information Administration (\texttt{InfoElectric@eia.gov}): \url{https://www.eia.gov/electricity/data/state/}}. As displayed, this is a time series $\Bp_t=[p_{t,1}\,p_{t,2}\, p_{t,3}]^\tr$ on the standard simplex $C^2$, so $p_{t,1}+p_{t,2}+p_{t,3}=1$ at all times. As such, if we define $u_{t,i}=\sqrt{p_{t,i}}$, then the transformed time series $\Bu_t=[u_{t,1}\,u_{t,2}\, u_{t,3}]^\tr$ is on the sphere $S^2$. \cite{sw2011jrssb} suggest that compositional data like these are best modeled on this spherical scale to avoid the problems associated with the more common log-ratio transformation for compositional data \citep{aitchison1982jrssb}. Our PDLM, the vMF-SSM, and the SAR are among the very few data-coherent time series models that generalize to this spherical case. These transformed compositional time series provide an opportunity to evaluate their performance beyond circular data. In the case of the SAR, we use the differenced SAR (DSAR) model for this energy shares application, just as \cite{zm2023joe} did in their original work. 

Table~\ref{tab:energy} displays one-step-ahead forecast metrics averaged over the period 2000 to 2022. Here, the DSAR generates the best point forecasts, with the PDLM not far behind, and the vMF-SSM twice as bad as the PDLM. In this small sample, the 90\% forecast sets from the vMF-SSM and PDLM both achieve 100\% empirical coverage, but the PDLM does so with significantly smaller sets. For density prediction, the PDLM generates a much smaller average kernel score than the vMF-SSM. Again, the DSAR does not generate set or density predictions. So overall, among the small basket of methods properly suited to this task, our PDLM provides competitive point predictions, and the only viable option in terms of set and density prediction.

\begin{table}
    \centering
    \begin{tabular}{l | llll}\hline
     & MSpFE  &  size  & coverage  & MKS  \\\hline 
DSAR     & 0.025 & -- &  -- & -- \\
vMF-SSM & 0.064 & 9.448 &  1.000 & -0.213 \\
PDLM & 0.033 & 0.166 &  1.000 & -0.474 \\
     \hline
\end{tabular}
    \caption{One-year-ahead, out-of-sample forecasting metrics for the energy shares data, averaged over the years 2000 to 2022. Forecast sets are $90\%$ posterior predictive credible sets.}
    \label{tab:energy}
\end{table}

\section{Conclusion}

There is clear empirical evidence (Figure~\ref{fig:results}, Table~\ref{tab:results}, etc.) that naive application of ordinary linear, Gaussian methods for spherical time series is strongly inadvisable. We can greatly improve predictive accuracy by using models that are tailored to data on $S^{n-1}$. There have been several such models proposed for the special case of $n=2$, but very few for the more general case of $n>2$, and little of this work has been from the Bayesian point of view. In this paper we proposed a state space model based on the projected normal distribution that can model spherical time series of arbitrary dimension. We described how to perform fully Bayesian inference in this model using an MCMC algorithm for offline inference and an RBPF for online inference. We also enumerated how to produce and evaluate the full range of probabilistic forecasts on a hypersphere of arbitrary dimension, which to the best of our knowledge is a first in this literature. In applications to forecasting wind direction and energy market composition, we showed that the proposed model outperforms its competitors in terms of point, set, and density prediction. 

There are numerous promising avenues for future work. First, we focused on the ``local level'' ($\BF_t=\BI_n$) variant of our model. Future work could explore other model variants, such as the dynamic regression model ($\BF_t=\BI_n\otimes \Bx_t^\tr$). Second, the proposed approach for set prediction applies broadly for predictive models on the $n$-sphere, including spatial models and regression models, and beyond the circular case of $n=2$. Thus, our methods may help construct interpretable summaries of these predictive models. Finally, with a few exceptions, spherical time series methods are notably deficient in software availability. Future work will build upon our replication files (\href{https://github.com/johnczito/ProjectedDLM}{\texttt{johnczito/ProjectedDLM}}) to provide user-friendly and freely-available statistical software for Bayesian time series on the $n$-sphere.

\bibliographystyle{ecta}
\bibliography{zito_references.bib}


\appendix 

\section{Calculating the trend in the local-level PDLM}
\label{app:trend}

Figure~\ref{fig:results} displays an estimate of ``trend'' for the local-level version ($\BF_t=\BI_n$) of the PDLM. This is the mean direction $\Bm_t=E(\Bu_t\given\Bs_t\com\BSigma)/||E(\Bu_t\given\Bs_t\com\BSigma)||_2$, which is \emph{not} equal to $\Bs_t$. It is a posterior functional of $\Bs_t$ and $\BSigma$ that is difficult to calculate exactly in the general case \citep{wg2013sm}. As such, we approximate the functional using Monte Carlo integration, as described in Algorithm~\ref{alg:trend}. Given posterior samples of $\Bs_t$ and $\BSigma$, we can apply Algorithm~\ref{alg:trend} on a draw-by-draw basis to produce posterior samples of the mean direction $\Bm_t$. Figure~\ref{fig:results} displays summaries of these draws.

\begin{algorithm}
\caption{Approximate the mean direction $E(\Bu)/||E(\Bu)||_2$ of $\Bu\sim\text{PN}_n(\Bs\com\BSigma)$}
\label{alg:trend}
\textbf{Input}: $\Bs$, $\BSigma$, $L$\;
Simulate $\tilde{\Bu}_1\com\tilde{\Bu}_2\com...\com\tilde{\Bu}_L\iid\text{PN}_n(\Bs\com\BSigma)$\;
Calculate $\bar{\Bu}=\sum_{l=1}^L\tilde{\Bu}_l/L$\;
\textbf{Return}: $\bar{\Bu}/||\bar{\Bu}||_2$.\\
(\textbf{Option}: if $n=2$, return $\text{atan2}(\bar{\Bu})\text{ mod }2\pi$)
\end{algorithm}

\section{Simulation: Algorithm~\ref{alg:pdlm_gibbs_sampler} ``gets it right''}
\label{app:gir}

In order to provide evidence that Algorithm~\ref{alg:pdlm_gibbs_sampler} has the correct invariant distribution and has been implemented correctly in the computer, we apply the check from \cite{geweke2004jasa}. The main idea behind \cite{geweke2004jasa} is that our Bayesian model specifies a joint distribution for both observables and unobservables:
\begin{equation}\label{eq:bayes}
p(\Bu_{1:T}\com r_{1:T}\com \state_{0:T}\com\Btheta)=\underbrace{p(\Bu_{1:T}\com r_{1:T}\given\state_{0:T}\com\Btheta)}_{(\ref{eq:measurement2})}\underbrace{p(\state_{0:T}\given\Btheta)}_{(\ref{eq:transition}\com \ref{eq:initialization1})}\underbrace{p(\Btheta)}_{(\ref{eq:gw}\com\ref{eq:sig})}.
\end{equation}
This joint distribution can be simulated in two ways:
\begin{itemize}
    \item (\textbf{marginal-conditional sampler}) first sample $p(\Btheta)$ from (\ref{eq:gw}\com\ref{eq:sig}), then $p(\state_{0:T}\given\Btheta)$ from (\ref{eq:transition}\com \ref{eq:initialization1}), then $p(\Bu_{1:T}\com r_{1:T}\given\state_{0:T}\com\Btheta)$ from (\ref{eq:measurement2}). This is a direct sampler that generates iid draws from the joint distribution;
    \item (\textbf{successive-conditional sampler}) alternate between sampling the conditional distributions $p(r_{1:T}\com\Bs_{0:T}\com\Btheta\given\Bu_{1:T})$ and $p(\Bu_{1:T}\given r_{1:T}\com\Bs_{0:T}\com\Btheta)$. This is a Gibbs sampler that simulates a Markov chain that has (\ref{eq:bayes}) as its stationary distribution. The first conditional is the usual posterior distribution that is targeted by our MCMC algorithm. The second conditional is the model likelihood (which is implicit in our case).
\end{itemize}
The marginal-conditional sampler is straightforward to implement using standard software. The successive-conditional sampler depends on the implementation of our MCMC algorithm. If everything (including the MCMC algorithm) has been implemented correctly, then the two samplers should generate draws from the same distribution, and we can verify this using probability plots or a statistical test. Furthermore, if we can conclude that Algorithm~\ref{alg:pdlm_gibbs_sampler} has been implemented correctly, we can directly compare its output to that of Algorithm~\ref{alg:particle_filter} (as we did in Figure~\ref{fig:correct}), which provides evidence that our particle filter has also been implemented correctly.

\begin{figure}
    \centering
    \includegraphics[width = 0.95\textwidth]{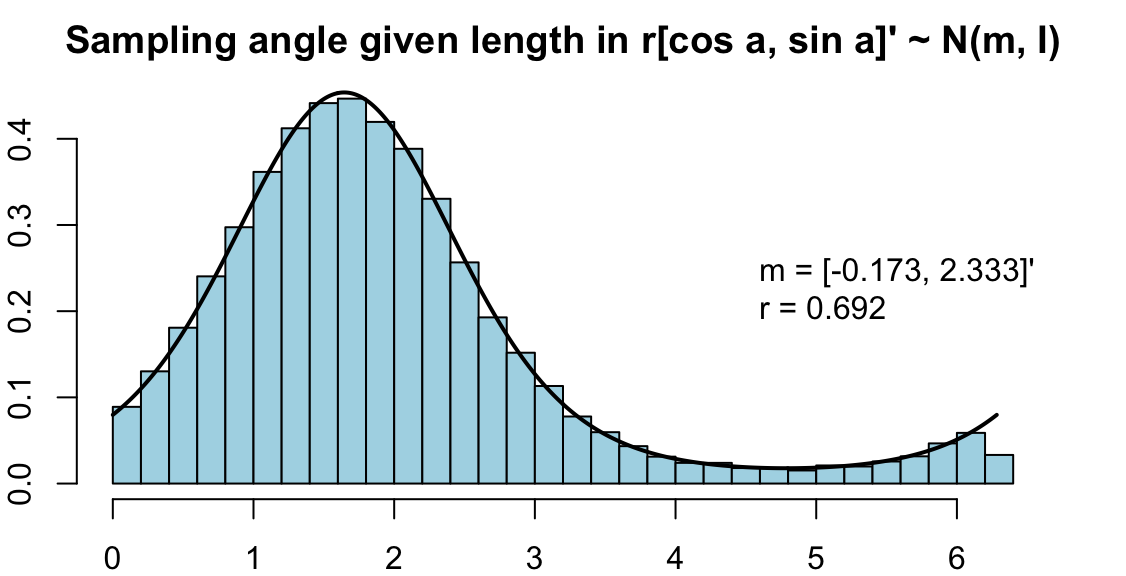}
    \caption{A histogram of draws from Algorithm~\ref{alg:AR} agrees with the density in (\ref{eq:dirpdf}). }
    \label{fig:AR}
\end{figure}

Implementing this check requires that we sample from $p(\Bu_{1:T}\given r_{1:T}\com\Bs_{0:T}\com\Btheta)$, which is difficult in general. However, in the special case with $n=2$ and $\BSigma =\BI_2$, it is possible. First, we reparametrize the unit vector observation $\Bu_t$ as $\Bu_t=[\cos a_t\,\sin a_t]^\tr$ for some $a_t\in[0\com2\pi)$, and we set $\Bmu_t=\BF_t\state_t$. We see in (\ref{eq:measurement2}) that the measurement distribution of the data-augmented state-space model is then
$$
r_t\Bu_t=r_t\begin{bmatrix}
\cos a_t \\ \sin a_t
\end{bmatrix}
\sim\textrm{N}_2\left(\Bmu_t=\begin{bmatrix}\mu_t^{(1)}\\\mu_t^{(2)}\end{bmatrix}\com\BI_2\right).
$$
Applying a change-of-variables to polar coordinates, we can rewrite the measurement density as 
$$
p(r_t\com  a_t\given \Bmu_t)
=
\frac{r_t}{2\pi}
\exp\left(
-\frac{1}{2}
\left[
\left(r_t\cos a_t-\mu_t^{(1)}\right)^2
+
\left(r_t\sin a_t-\mu_t^{(2)}\right)^2
\right]
\right).
$$
So the conditional density for $ a_t$ is
\begin{equation}\label{eq:dirpdf}
    p(a_t\given r_t\com \Bmu_t)
=
\frac{p(r_t\com a_t\given\Bmu_t)}{\int_0^{2\pi} p(r_t\com a_t\given\Bmu_t)\,\dd  a_t}
=
\frac
{
\exp
\left(
r_t
\mu_t^{(1)}
\cos a_t
+
r_t
\mu_t^{(2)}
\sin a_t
\right)
}
{2\pi I_0(r_t||\Bmu_t||_2)},
\end{equation}
where $I_0$ is the modified Bessel function of order 0.
When $r_t=1$ and $\Bmu_t=[\cos\lambda\,\sin\lambda]^\tr$ for some $\lambda\in[0\com2\pi)$, then this is the density of the von Mises distribution with location paramater $\lambda$ and concentration parameter $\kappa=1$. We can show that 
$$
M = \max_{0< a_t<2\pi}\frac{p( a_t\given r_t\com\Bmu_t)}{1/2\pi}
=
\frac
{
\exp
\left(
r_t
||\Bmu_t||_2
\right)
}
{I_0(r_t||\Bmu_t||_2)}
,
$$
and this enables us to implement Algorithm~\ref{alg:AR} to sample directly from $p(a_t\given r_t\com\Bmu_t)$. Figure~\ref{fig:AR} displays a histogram of draws from Algorithm~\ref{alg:AR} and a line plot of the density in (\ref{eq:dirpdf}), and they clearly agree. Because the $\Bu_{1:T}$ are conditionally independent given the $\Bs_{0:T}$ and $r_{1:T}$, we can sample from $p(\Bu_{1:T}\given r_{1:T}\com\Bs_{0:T}\com\Btheta)$ by separately applying Algorithm~\ref{alg:AR} for each $t$.

\begin{algorithm}
\caption{Accept-reject sampler for $p(a\given r\com\Bmu)$ when $\begin{bmatrix}r\cos a & r\sin a\end{bmatrix}^\tr\sim\text{N}_2(\Bmu\com\BI_2)$}
\label{alg:AR}
\textbf{Input}: $r$, $\Bmu$\;
Calculate $M=\exp
\left(
r
||\Bmu||_2
\right)
/I_0(r||\Bmu||_2)$\;
Draw $U\sim\textrm{Unif}(0\com 1)$\;
Propose $V\sim\textrm{Unif}(0\com 2\pi)$\;
\While{$U\geq2\pi p(V\given r\com\Bmu)/M$}
{
Draw $U\sim\textrm{Unif}(0\com 1)$\;
Propose $V\sim\textrm{Unif}(0\com 2\pi)$\;
}
\textbf{Return}: $V$.
\end{algorithm}

\begin{figure}
    \centering
    \includegraphics[width = 0.9\textwidth]{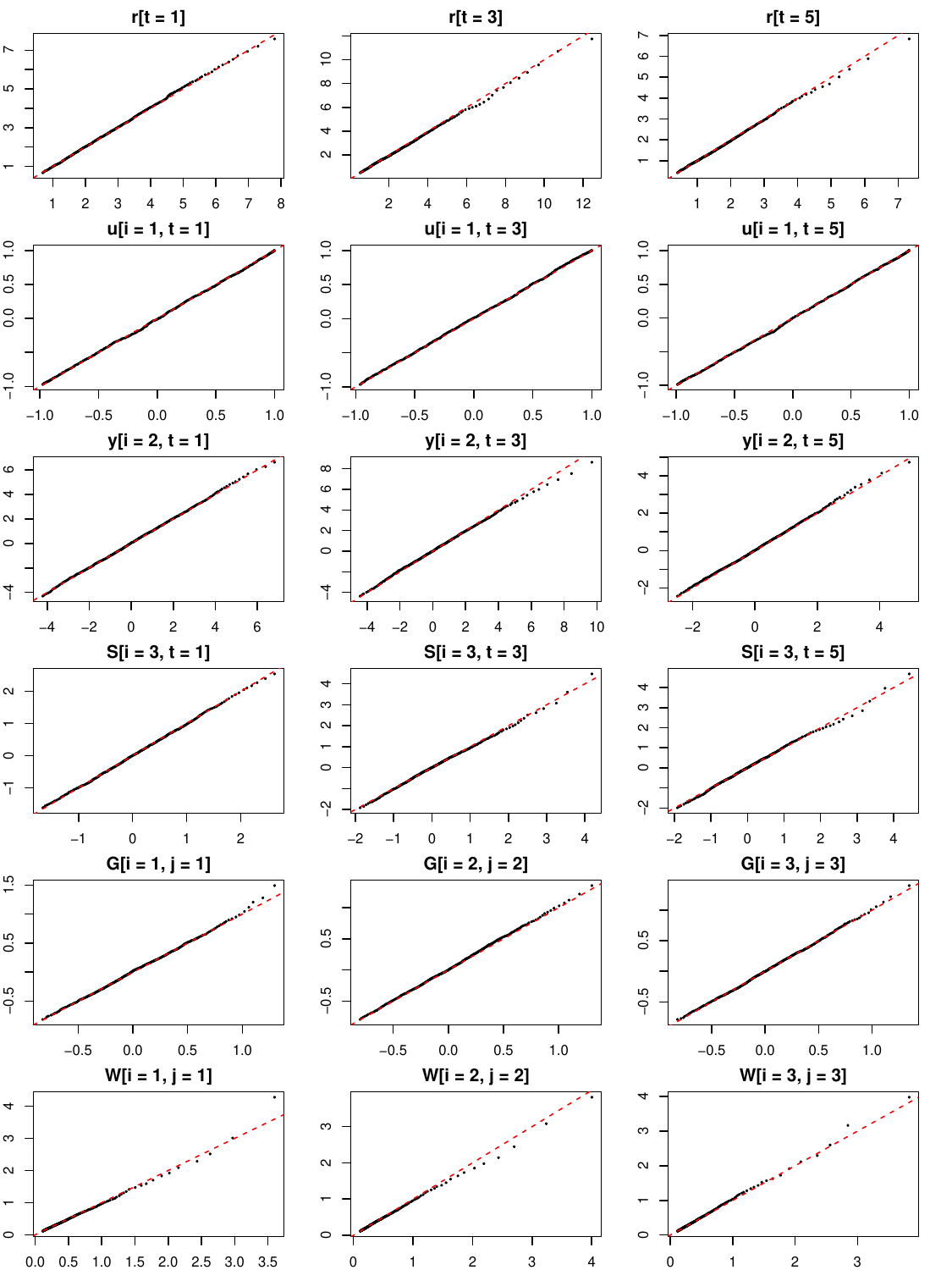}
    \caption{Comparing quantiles of the marginal-conditional draws (horizontal axis) and successive-conditional draws (vertical axis) for various elements of $(\Bu_{1:T}\com r_{1:T}\com \Bs_{1:T}\com\Btheta)$. If the quantiles agree, this provides evidence that the two samplers generate draws from the same distribution, which should be the case if our MCMC algorithm is error-free.}
    \label{fig:gir}
\end{figure}

With this, we apply \cite{geweke2004jasa} with the following settings:
\begin{itemize}
    \item $n=2$; $p = 3$; $T = 5$; $\BSigma = \BI_n$; each $\BF_t$ filled with iid draws from $\text{N}(0\com 1)$;
    \item $\bar{\Bs}_{1|0}=\Bzero$; $\BP_{1|0}=\BI_p$; $\nu_0=p+2$; $\BPsi_0=\BI_p$; $\overline{\BG}_0=\Bzero$; $\BOmega^{-1}_0=\BI_p$;
    \item 5,000 draws from each sampler; the successive-conditional draws thinned down from 50,000.
\end{itemize}
Lastly, note that because we are fixing $\BSigma=\BI_n$, we are testing the special case of Algorithm~\ref{alg:pdlm_gibbs_sampler} that skips the steps for $\BGamma$ and $\Bgamma$. With that, Figure~\ref{fig:gir} displays Q-Q plots comparing the draws from the marginal-conditional sampler (horizontal axis) to the draws from the successive-conditional sampler (vertical axis) for a selection of observables and unobservables in the PDLM. In all cases, we see that the Q-Q points lie on the 45 degree line, which provides evidence that the two samplers are generating draws from the same distribution. This is what we would expect to see if all of our code (including our implementation of Algorithm~\ref{alg:pdlm_gibbs_sampler}) is error-free.








\section{Simulation: parameter recovery}
\label{app:consistency}

\begin{figure}
    \centering
    \includegraphics[width=0.85\textwidth]{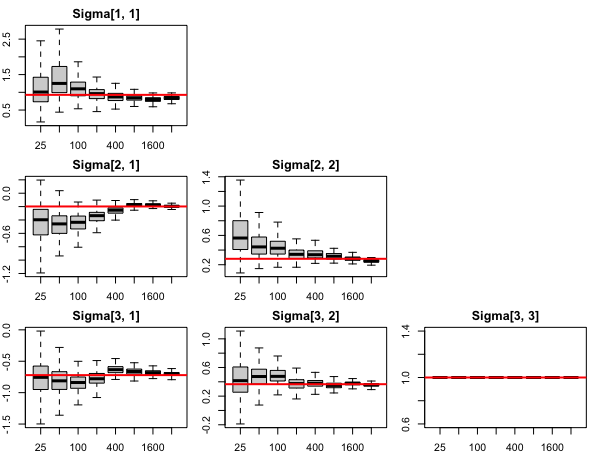}
    \includegraphics[width=0.85\textwidth]{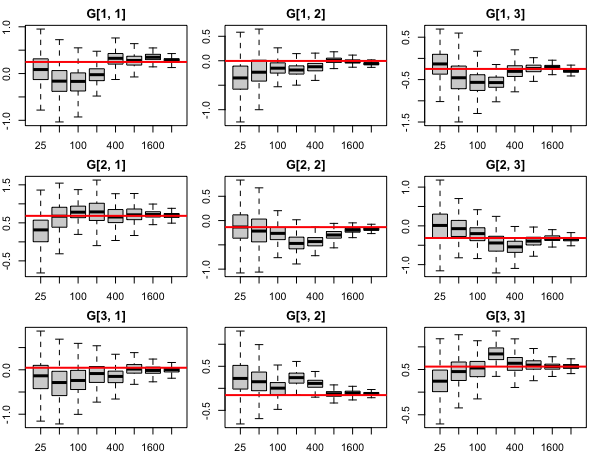}
    \caption{As we fit the PDLM to more and more simulated data, the posterior concentrates around the ground truth values for the static model parameters $\BSigma$ and $\BG$.}
    \label{fig:consistency}
\end{figure}

Appendix~\ref{app:gir} checked Algorithm~\ref{alg:pdlm_gibbs_sampler} in the case of $n=2$ and $\BSigma=\BI_n$. For a cruder check that encompasses the general case and also evaluates the statistical self-consistency of the PDLM, we perform a parameter recovery exercise. We set the following:
\begin{itemize}
    \item $n=3$; $p = n$; each $\BF_t$ filled with iid draws from $\text{N}(0\com 1)$;
    \item $\text{vec}(\BG^{(\text{true})})\sim\text{N}^\star_{p^2}\left(\text{vec}(0.5\cdot\BI_p)\com\BI_{p^2}\right)$; $\BGamma^{(\text{true})}\sim\text{IW}_{n-1}(n+1\com \BI_{n-1})$; $\Bgamma^{(\text{true})}\sim\text{N}_{n-1}(\Bzero\com\BI_{n-1})$; $\BW^{(\text{true})}\sim\text{IW}_p(p+2\com \BI_p)$;
    \item $\bar{\Bs}_{1|0}=\Bzero$; $\BP_{1|0}=\BI_p$; $\nu_0=p+2$; $\BPsi_0=\BI_p$; $\overline{\BG}_0=\Bzero$; $\BOmega^{-1}_0=\BI_p$; $d_0=n+1$; $\BPhi_0=\BI_{n-1}$; $\bar{\Bgamma}_0=\Bzero$; $\BLambda_0=\BI_{n-1}$.
\end{itemize}
Based on the randomly-generated ground truth parameter values, we simulate a long time series of length $T=3200$ from the model. We recursively refit the model on an expanding window of this synthetic data, and observe in Figure~\ref{fig:consistency} that the posterior distribution (visualized using boxplots of the MCMC draws) concentrates around the ground truth values (red lines) as the sample size grows.

\end{document}